\documentclass{sig-alternate} 
\usepackage{tikz,multirow,comment,float,makecell,etoolbox,amsmath,amssymb,graphicx,rotating,color,soul,cite,threeparttable,colortbl,xcolor,flushend,fancyhdr,mathptmx}
\usepackage[english]{babel}
\graphicspath{{fig/}}

\usepackage{mathptmx} 

\newcommand{\ignore}[1]{}
\usepackage[normalem]{ulem}
\usepackage[hyphens]{url}
\usepackage{microtype}
\usepackage[bookmarks=true,breaklinks=true,letterpaper=true,colorlinks,linkcolor=black,citecolor=blue,urlcolor=black]{hyperref}

\newcommand{\iscasubmissionnumber}{847}

\fancypagestyle{firstpage}{
  \fancyhf{}

  \fancyhead[C]{\normalsize{ISCA 2019 Submission
      \textbf{\#\iscasubmissionnumber} \\ Confidential Draft: DO NOT DISTRIBUTE}} 
  \fancyfoot[C]{\thepage}
}  

\newcommand{\squishlist}{
   \begin{list}{$\bullet$}
    { \setlength{\itemsep}{0pt}      \setlength{\parsep}{0pt}
      \setlength{\topsep}{3pt}       \setlength{\partopsep}{0pt}
      \setlength{\listparindent}{-2pt}
      \setlength{\itemindent}{-5pt}
      \setlength{\leftmargin}{1em} \setlength{\labelwidth}{0em}
      \setlength{\labelsep}{0.5em} } }

\newcommand{\squishend}{
    \end{list}  }

\newcommand{\Fsize}[2][.]{%
    \ifthenelse{\equal{#1}{.}}{
        \fontsize{#2}{#2}%
    }{%
        \fontsize{#2}{#1}%
    }%
    \selectfont%
}

\definecolor{LightCyan}{rgb}{0.88,1,1}
\setstcolor{gray}
\newcommand*\circled[1]{\tikz[baseline=(char.base)]{
            \node[shape=circle,draw,inner sep=0.6pt,text=white,fill=black] (char) {#1};}}

\newcommand{\xname}{MEDAL\xspace}
\newcommand{\other}{\emph{et al.}\xspace}

\newcommand{\todo}[1]{{\hl{[Question]: #1}}}%

\soulregister\other7
\soulregister\circled7
\soulregister\cite7
\soulregister\ref7
\soulregister\bf7
\soulregister\small7
\soulregister\emph7
\soulregister\xname7
\soulregister\textbf7
\soulregister\textit7

\makeatletter
\newcommand{\setword}[2]{%
  \phantomsection
  #1\def\@currentlabel{\unexpanded{#1}}\label{#2}%
}
\makeatother

\renewcommand{\paragraph}[1]{\noindent\textbf{\underline{#1:}\xspace}}

\excludecomment{Hidden}
\newcommand{\Hide}[1]{}

\specialcomment{Notes}{%
    \noindent$\Rightarrow\Rightarrow$%
}{%
    $\Leftarrow\Leftarrow$%
}

\newcolumntype{I}{!{\vrule width 0.8pt}}

\usepackage[normalem]{ulem}
\usepackage{cite}

\begin{document} 
\title{Neural Network Model Extraction Attacks in Edge Devices\\by Hearing Architectural Hints\vspace{-50pt}}
\author{Xing Hu$^1$, Ling Liang$^1$, Lei Deng$^1$, Shuangchen Li$^1$, Xinfeng Xie$^1$, Yu Ji$^1$, \\Yufei Ding$^1$, Chang Liu$^2$, Timothy Sherwood$^1$, Yuan Xie$^1$\\
$^1$University of California, Santa Barbara,
$^2$University of California, Berkerly\\
\{huxing, lingliang, leideng, shuangchen, xinfengxie, yuji, yuanxie\}@ucsb.edu, \\ \{yufeiding,sherwood\}@cs.ucsb.edu, liuchang@eecs.berkeley.edu\vspace{50pt}}

\date{}
\maketitle
\pagestyle{plain}


\newif\ifversion
\versiontrue

\begin{abstract}

As neural networks continue their reach into nearly every aspect of software operations, the details of those networks become an increasingly sensitive subject.   
Even those that deploy neural networks embedded in physical devices may wish to keep the inner working of their designs hidden -- either to protect their intellectual property or as a form of protection from adversarial inputs.  
Prior work has demonstrated that image classifiers can be attacked with nearly imperceptible malicious perturbations to intentionally cause misclassification of the recognized outputs and that details of the network structure are critical to carrying out such attacks with high probability.  
We show that physical deployments of these networks running on microprocessors and GPUs unfortunately shed incredibly detailed information about internal network structure on easily probed interfaces.

The specific problem we address is how, given noisy and imperfect measurements of the off-chip memory buses, one might reconstruct the the neural network architecture including the set of layers employed (e.g. ReLU, Convolution, etc), their connectivity, and their respective dimension sizes.  
Considering both the intra-layer architecture features and the inter-layer temporal association information introduced by the DNN design empirical experience, we draw upon ideas from speech recognition to solve this problem. We show that off-chip memory address traces and PCIe events provide ample information to reconstruct such neural network architectures accurately.
We demonstrate that these concepts work experimentally in the context of an off-the-shelf Nvidia GPU platform running a variety of complex neural network configurations, 
and that the result of this reverse engineering effort is directly helpful to those attempting to craft adversarial inputs.  
In the end, our techniques achieve a high reverse engineering accuracy and improve the one's ability to conduct targeted adversarial attack with success rate from 14.6\%$\sim$25.5\% (without network architecture knowledge)  to 75.9\% (with extracted network architecture).

\end{abstract}
\section{Introduction}\label{sec:introduction}


Machine learning approaches, especially deep neural networks (DNNs), are transforming a wide range of application domains, such as computer vision~\cite{NIPS2012:Hinton,VGG2014,he2016deep}, speech recognition \cite{xiong2017microsoft}, and language processing~\cite{NIPS2014:RNN, ICML2008:NLPNN, vaswani2017attention}. Computer vision in particular has seen commercial adoption of DNNs with impacts across the automotive industry, business service, consumer market, agriculture, government sector, and so forth~\cite{tractica_report}. For example, autonomous driving, 
with \$77 billion projected in revenue by 2035, has attracted the attention of giants including Tesla, Audi, and Waymo~\cite{tesla_news, tesla_news2,audi_nvidia,waymo1}. Despite the rising opportunities for DNNs to benefit our life~\cite{ucb},  the security problems introduced by DNN systems have emerged as \textbf{an urgent and severe problem}, especially for mission critical applications~\cite{atm_security,Ahmed_2015_CVPR,NNsafetysummray:papernot2016,adversary:survey}.

As DNN models become more important in system design, protecting DNN model architecture information becomes more critical both due to security concerns and intellectual property protection. Black-box DNNs that encapsulate the internal model characteristics information, have become the mainstream in the AI development community. By extracting the model information, attackers can not only counterfeit the intellectual property of the DNN design, but also conduct more efficient adversarial attacks towards the DNN system ~\cite{ICLR17:Delving:Dawn,meta_learning}.
{The commonly-used deployment strategy,
``Cloud Training Edge Inference'', makes the model extraction more destructive.  It is appealing for attackers
to do the physical inception on edge devices, because the success of hacking one device can be leveraged to unlock many
other devices sharing the same neural network model.} 


The AI community envisions the importance of the neural network security and abundant studies come out. Prior studies mainly conduct model extraction through detecting the decision boundary of the victim black-box DNN model~\cite{fgsm,miFGSM}. Nevertheless, such approach demands significant computational resources and huge time overhead: given the pre-knowledge of the total number of layers and their type information, it still takes 40 GPU-days to search a 7-layer network architecture with a simple, chain topology~\cite{meta_learning}. Even worse, this approach cannot accommodate state-of-the-art DNNs with complex topology, e.g., DenseNet~\cite{densenet} and ResNet~\cite{resnet}, due to the enlarged search space of possible network architectures.

Real attackers actually have more information at their disposal than might be suggested by algorithm-centric attacks.  We show that a more systematic approach, using architecturally visible information 
has a great deal of power. A complete DNN system includes several components: the DNN model, the software framework, and the hardware platform. These components are not independent of each other. 
Information leakage from the hardware platform, for example, could expose the kernel events in the DNN model execution and potentially unveil the entire DNN model.

Unlike the attacks on accelerators~\cite{DAC18:ReverseNN}, attacking on GPU platform are much challenging because of the heavier system stack and deeper memory hierarchy.
With the system stack, DNN layers are transformed into many GPU-kernels dynamically during run-time(Figure.3). E.g., A single CONV can end up with >10 GPU-kernels under different implementations(e.g. Winograd/Fourier,etc.); then with 65-1500 kernels for a typical DNN, it is difficult to even figure out layer number/boundary, not to mention their structure/connection. Accelerator attacks~\cite{DAC18:ReverseNN} do not consider such problem.
Additionally, the unique comprehensive memory optimization on GPU raise the difficulty, because of: (1) unknown address mapping from logic address to physical address, and then to device address; (2) noisy memory traffic for intermediate data(For example, the data with read-only access may come from workspace in cuDNN); (3) incomplete memory accesses due to optimization for data reuse and computing parallelism. Therefore, it is extremely challenging to  accurately identify the layer sequence based on the imperfect execution statistics (with filtering and run-time scheduling) of very long kernel sequence (10x~1000x). 
 

To address these issues, we propose a methodology which extracts models fully exploiting both architecture execution features visible off-chip (e.g. memory access behavior) and priors learned from the rich families of DNN now in operation. 
Considering both the intra-layer architecture execution features and inter-layer temporal association likelihood, we draw upon ideas from speech recognition to achieve accurate model extraction. 
A central idea of the paper is that inter-layer DNN architecture features can be considered ``sentences'' in a language of DNNs.  Just like natural language our reading of any individual ``word'' may be quite error prone, but when placed into the context of the ``sentence'' we can find a parsing that maximizes the likelihood of a correct match 
far more effectively than character-by-character approaches ever could. We show that off-chip memory address traces and PCIe events provide ample information to reconstruct neural network architecture accurately with this greater context.
In a summay, \textbf{we are the first to propose and demonstrate end-to-end attacks in the context of an off-the-shelf Nvidia GPU platform with full system stack}, which urges the demand to design secure architecture and system to protect the DNN security . 

In summary, we make the following contributions:

\squishlist

  \item We propose a holistic method which considers both the intra-layer architecture features and inter-layer temporal association likelihood introduced by the DNN design empirical experience to conduct accurate model extraction.
  We show that off-chip memory address traces and PCIe events provide ample information to reconstruct neural network architecture accurately.
  \item We formalize the neural network architecture extraction as a sequence prediction problem, and solve this problem with a sequence model using analogous speach recognition techniques that achieves high accuracy and generality. Building upon this information, we show how one can reconstruct the  layer topology and explore dimension space with the assistance of the memory bus traffic information, and finally form the complete neural network architecture.
  \item We experimentally demonstrate our methodologies on an off-the-shelf GPU platform. With the easy-to-get off-chip bus communication information, the extracted network architectures exhibit very small difference from that of the victim DNN models.
  \item We conduct an end-to-end attack to show that the extracted model boosts the attacking effectiveness of adversarial attack, which introduces 50.4\% improvement of attacking success rate compared to cases without neural network architecture knowledge. We \emph{demonstrate} that memory address traces are able to damage the NN system security which urges hardware security studies(e.g.ORAM), raising the attention of the architecture/system community to build more robust NN system stack
  
\squishend

\section{Background and Motivation}

In this section, we introduce the background of model extraction and existing model extraction techniques. 


\subsection{Model Characteristics}

Model extraction attacks aim to explore the model characteristics of DNNs for establishing a near-equivalent DNN model~\cite{stealing_para}.
It is the initial step for further attack. For example, to attack a victim black-box model, the adversary needs to build a substitute model for generating adversarial examples~\cite{fgsm,miFGSM,NNsafetysummray:papernot2016,adversary:survey}, while the similarity of  characteristics between substitute and victim models strongly impacts the effectiveness of these adversarial examples~\cite{ICLR17:Delving:Dawn,meta_learning}.

The model characteristics one would hope extract include: (1) \textbf{\emph{network architecture}} consists of layer depth and types, connection topology between layers, and layer dimensions (including channel number, feature map and weight kernel size, stride, and padding etc). 
(2) \textbf{\emph{parameters}} include the weights, biases, and Batch Normalization (BN) parameters. They are updated during stochastic gradient descent (SGD) in the training process.
(3) \textbf{\emph{hyper-parameters}} include the learning rate, regularization factors, and momentum coefficients, etc. The hyper-parameters are statically configured at the beginning and will not be updated during SGD. 

{Among all of the model characteristics, the network architecture is most fundamental for NN security}. The model parameters, hyper-parameters, and even training data may be inferred with the knowledge of the network architecture~\cite{stealing_para,stealing_hyperpara}. Moreover, previous work~\cite{ICLR17:Delving:Dawn,meta_learning} observes that the network architecture similarity between the substitute and victim model plays a key role for attack success rate.


\subsection{Algorithm vs. Holistic Approaches}

\begin{figure}[!htbp]
  \centering
  \includegraphics[width=\linewidth]{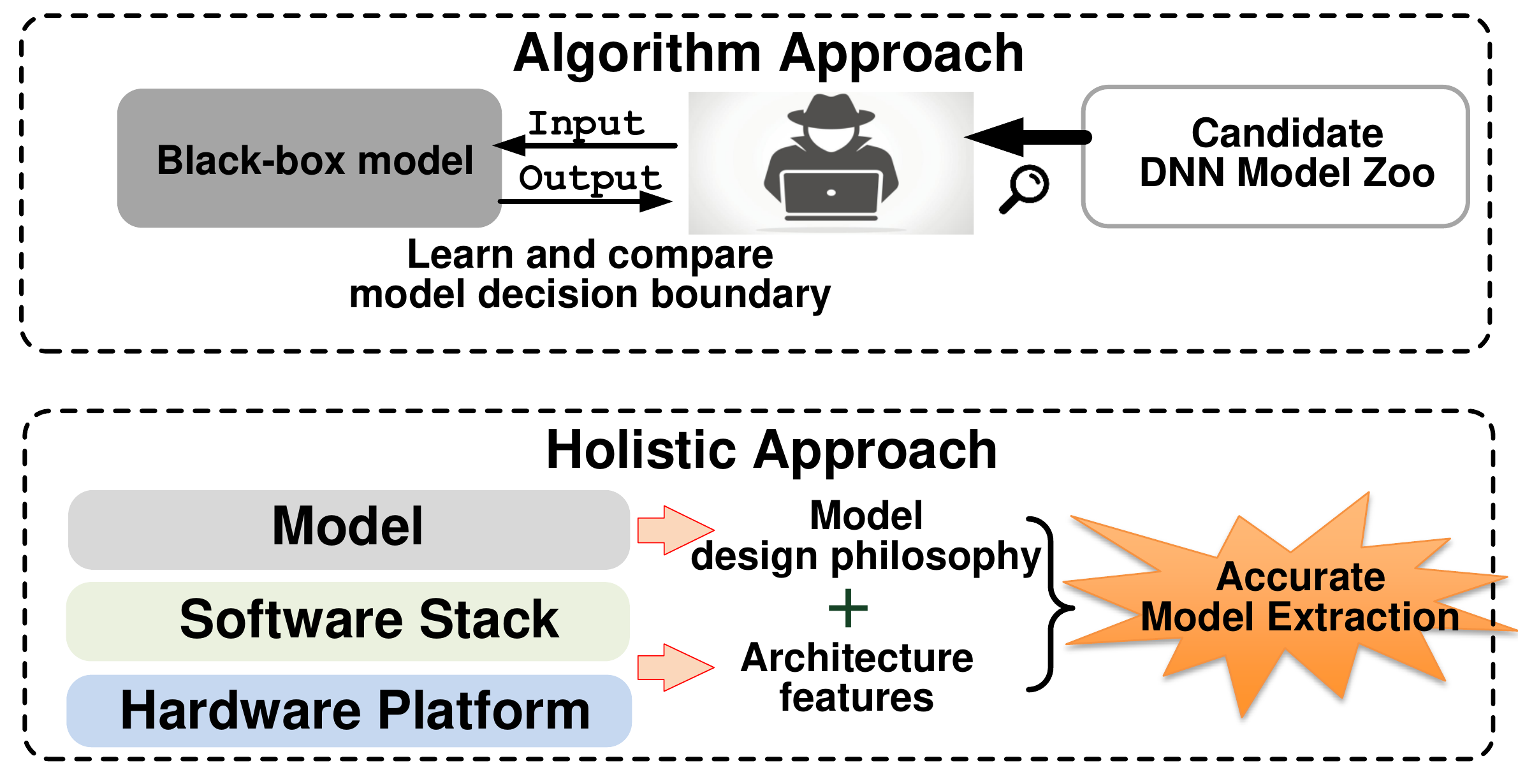}
  \vspace{-15pt}
 \caption{(a) Prior approach merely relying on  algorithm; (b) Proposed holistic approach.}\label{fig:algo_vs_holis}
\end{figure}

Many algorithms designed for model extraction have been proposed~\cite{meta_learning,stealing_para}. Unfortunately, they require the prior knowledge of the network architectures and significant computation demands. As shown in Figure~\ref{fig:algo_vs_holis}, the key idea of an algorithm-centric approach is to search the models in the candidate model zoo to find the one with closest decision boundary as the victim model. The models in the candidate model zoo are trained with the input and output obtained by querying the victim model. 
However, it is extremely challenging to apply this type of method to extract complex DNN models. Two different networks may have the similar input--output responses for most queries, which makes such methods inherently inaccurate. Furthermore, unlike the parameters, the network architecture cannot evolve dynamically during the learning process. The result of these challenges is that we need to enumerate all possible network architectures to find the closest one, which consumes significant computation resources. It is almost impossible to find the victim model which hasn't been released before. 


 We explore the opportunity for architecture execution features to be exploited to help achieve better model extraction in another perspective, as shown in Figure~\ref{fig:algo_vs_holis}. 
 Although prior studies starts to consider the potentiality of leveraging architecture information leakage~\cite{DAC18:ReverseNN}, prior work focuses on the customized FPGA DNN accelerator which have less complex system dynamics  which then results in a system easier to reverse engineer.
 In this work, we demonstrate end-to-end attacks extractions in the context of GPUs which pull from a richer set of possible layers and suffer from much noisier architectural measurements. When we consider both layer architecture features and the inter-layer association probabilities, we show that it is still possible to conduct accurate model extraction even with the architecture and system noises. 



\section{Attack Overview}\label{sec:attack_flow}

In this section, we introduce the threat model and the specific hardware information obtained during execution.   

\subsection{Attack Model}
\begin{figure}[!htbp]
  \centering
  \includegraphics[width=\linewidth]{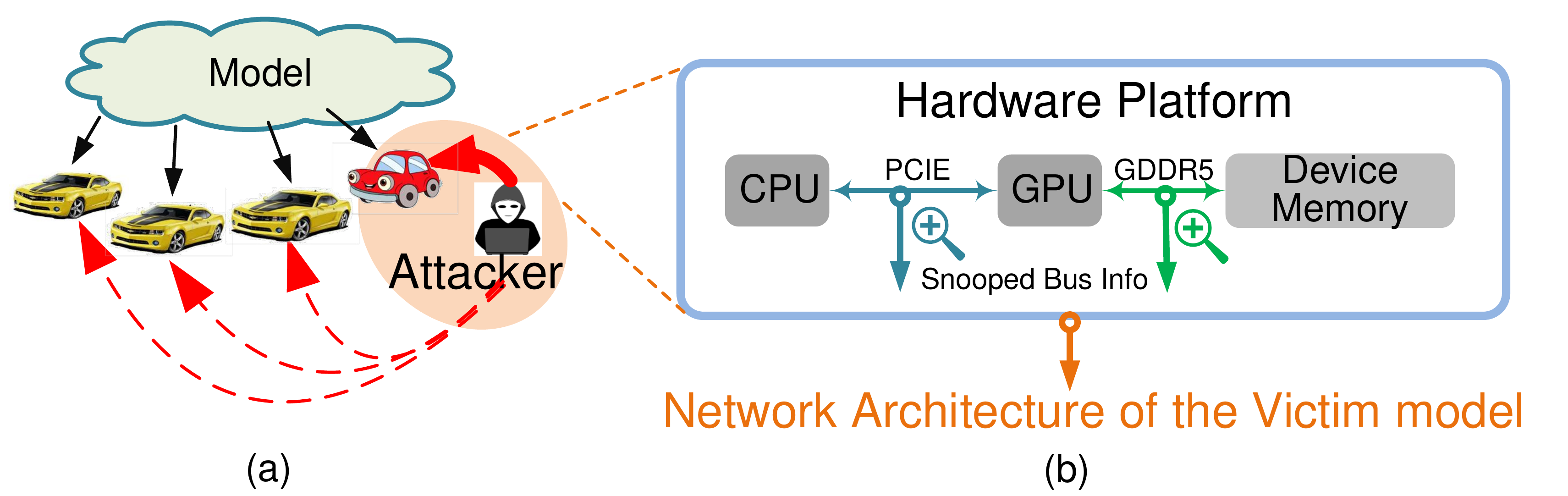}
  \caption{Illustration of the attack model. (a) Hack-one, Attack-All-Others. (b). Bus snooping at GPU platform. }\label{fig:Main_idea}
\end{figure}

We focus on the edge security in this work. As shown in Figure~\ref{fig:Main_idea}, the attacker can physically access one edge device encapsulating a victim DNN model for model extraction and attack all the other devices which share the same neural network model.
We assume that the adversary use bus snooping techniques which passively monitor PCIe and memory bus events.  {Bus snooping is a well understood, practical, and low-cost attack that has been widely demonstrated~\cite{physical_attack:2002,DMA_Attack:12,HMTT}}. We do not assume that the attacker has any access to the data passing through through buses, only the addresses, and the attacks described here can work even when data is encrypted. 
We make no assumptions with regard the ability of the attacker to know even what family of DNN models might be running, what software codes might implement those models, or have any other information about the operation of the device under attack that is not directly exposed through externally accessible connections.  The model extraction parts of the attack are fully passive requiring only the ability to observe architectural side channels over time.  To complete the attack and craft adversarial inputs, the ability to provide specific inputs and observe results is also required.

\subsection{Target Hardware Platform} 

To make the attack more concrete, we specifically consider a heterogeneous CPU-GPU platform. 
The basic infrastructure~\cite{CDMA} is shown in Figure~\ref{fig:Main_idea}.(b). The CPU and GPU are connected by the PCIE bus, and the host and device memories are attached to the CPU and GPU through DDR and GDDR memory buses, respectively. Such a design offers good programmability, generality, high performance, and hence is a representative platform for such attacks.
Many real industrial products are built around such an architecture including most of the existing L3 autopilot systems~\cite{waymo1, audi_nvidia}. The adversary can get access to the PCIe and GDDR bus for model extraction~\cite{physical_attack:2002,DMA_Attack:12,HMTT}, either by physical probing at the interconnect~\cite{physical_attack:2002} or applying a DMA capable device~\cite{DMA_Attack:12}.  

\subsection{Architecture Information Leakage} 
Table~\ref{tab:info_summary} lists the information we can get from the PCIe bus and device memory bus.

\begin{table}[!hbtp]\vspace{-5pt}
\caption{\label{tab:info_summary} Bus snooped information.} 
\small
\centering
\renewcommand\arraystretch{1}
\setlength{\tabcolsep}{5pt}
    	\begin{tabular}{|c|c|c|c|}
    	\hline
    	   & \textbf{Obtained} & \textbf{Inferred} \\ \hline
    	\textbf{PCIe Bus} &  Kernel events;  & $Exe_{Lat}$\\ 
    	& \emph{Mem copy size ($Mem_{cp}$)}& \\\hline
    	\textbf{Device Memory Bus} &  Memory  request trace & $R_v$, $W_v$, $d_{RAW}$\\ \hline
    \end{tabular}
\end{table}

\subsubsection{Information Leakage Through PCIe Bus.}~\label{sec:pcie_data}
\vspace{-10pt}

\noindent\textbf{\emph{Obtained Information:}}
According to the GPU programming model, the CPU transfers data from the host memory to the device memory and then launches GPU kernels for execution. Once the GPU finishes the task, it transfers results back to the host memory. Thus, there are copy events or control messages through the PCIe bus before the kernel launching and completion during the CUDA program execution~\cite{GPUkernel:13}. The attacker can obtain two kinds of information: the kernel events and the \texttt{memory copy size} ($Mem_{cp}$) between CPU and GPU. 

\noindent\textbf{\emph{Inferred Information:}}
From the data above we can infer the \texttt{kernel duration time} ($Exe_{lat}$) from the kernel events.

\subsubsection{Information Leakage Through  Memory Bus.} 
\noindent\textbf{\emph{Obtained Information:}} During the process of memory bus snooping, the memory access type (read or write), address, and a time stamp for each access can be obtained~\cite{HMTT}. 

\noindent\textbf{\emph{Inferred Information:}}
According to the time stamp of memory requests and the kernel execution period, we can infer the following architectural execution characteristics. (1) Read and write data volume ($R_v$ and $W_v$) of the memory requests in every kernel. (2) The data reuse kernel distance according to the addresses and types of memory requests. Specifically, we focus on reuse distance in the kernel wise of the Read after Write (RAW) pattern which is referred to as $d_{RAW}$. 


\section{Network Architecture Extraction}

\begin{figure}[!htbp]
  \centering
  \includegraphics[scale=0.30]{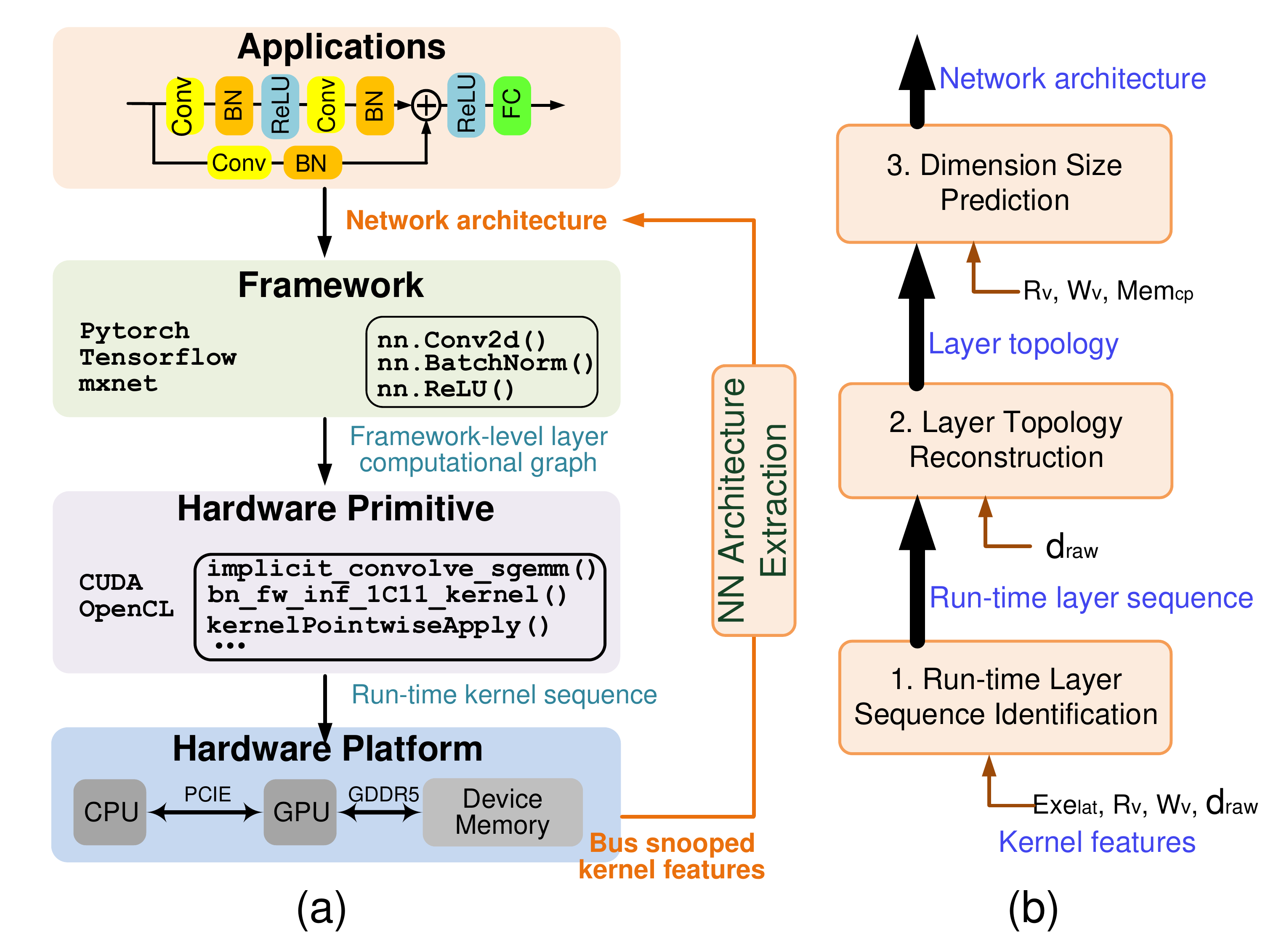}
  \vspace{-15pt}
 \caption{(a). DNN system stack. (b)Three-step methodology for network architecture extraction .}\label{fig:systemstack}
\end{figure}

At a high level, the goal of the proposed methodology is to leverage the hardware snooped information to extract the network architecture of victim model, including the set of layers employed, the connections between layers, and their respective dimension size. 
 This reverse engineering goes through multiple layers of the DNN system stack, including framework, primitive, and hardware platform.  
 As shown in Figure~\ref{fig:systemstack}.(a),   
 the frameworks optimize the network architecture to form the framework-level layer computational graph and transform these high-level abstractions to hardware primitives (cuDNN, OpenCL) for better resource utilization. The cuDNN library \cite{cuDNN} launches the 
 well-optimized handcraft kernel sequence according to the layer type. Finally,  kernel execution on the hardware platform exhibits architecture features, including the memory access pattern and the kernel execution latency.

There are several challenges to achieving the goal of extracting network architecture based on kernel execution feature sequences: 
1) The relationship between layer and kernel is not static one-to-one correspondence relationship. For example, single conv layer may be implemented as 10x-100x different kernels and in 7 different implementations during run-time. Therefore accurately identify the layer sequence based on the execution statistics of very long kernel sequence (10x~1000x) is an important and challenging task.
1) Some kernels belonging to different layers have quite similar architectural execution features, such as BN, ReLU, and some kernels from Conv.
2) Memory hierarchy and programming library optimization increase the variations of the architecture events, which introduces the run-time noises into the kernel execution features. For example, the cuDNN \cite{cuDNN} library greatly optimizes the convolution and matrix-vector multiplication. There are seven different algorithm implementations for the Conv layer, which are selected during running time, aiming at fully leveraging the compute capability of GPU resources for better performance. Hence, the Conv layers produce variable numbers of execution kernels with different features. Overall, these architectural and system designs introduce noises that lower the identification accuracy for recovering the DNN network structure.


To address these issues, we propose a methodology which employs both architecture execution features and inter-layer context probability of building DNN models. The overall process consists of three steps: 1) Run-time layer sequence identification; 2) Layer topology reconstruction; and 3) Dimension size estimation, as shown in Figure~\ref{fig:systemstack}.(b). The three steps of performing the model extract then are as follows:

\label{sec:architecture}


\subsection{Run-time Layer Sequence Identification}


In this step, the attacker identifies the executed layers during running time according to the kernel execution features. We first analyze the characteristics of different layers. We identify that both the kernel execution features and layer context features are important for run-time layer sequence identification. We then ingeniously formalize the layer sequence identification as a sequence-to-sequence problem. At the end, we leverage a speech recognition approach~\cite{ctc} as a tool to solve this problem, which achieves accurate prediction.

After comprehensively investigating modern DNN models, we consider the following layers in this work: Conv (convolution), FC (fully-connected), BN (batch normalization), ReLU (rectified linear unit), Pool, Add, and Concat, because most of the state-of-art neural network architectures can be represented by these basic layers~\cite{VGG2014,GoogleNet,resnet,polynet,nasnet,BENCHIP}. Note that it is easy to integrate other layers into this methodology if necessary. Every layer conducts a certain operation for the input data and output results to the next layer(s). They have following functionality: 1) Conv and FC implement linear transformations on the input or activation data. 2) ReLU performs nonlinear transformations on the input activation, which has equal input and output data volumes. 3) BN performs normalization (e.g. scaling and shifting) on the input activation for faster convergence and also has equal input and output volumes. 4) Pool aggregates features by down-sampling the input activation for dimension reduction.  5) Add performs element-wise addition on two input activation tensors.  6) Concat concatenates several sub-input tensors into a single output tensor \cite{GoogleNet}.
To identify which layer the kernels belong to, we first analyze the characteristics of these layers in terms of both {architectural behavior} and {model design principle}.

\subsubsection{Layer Characterization}

\noindent\textbf{\ul{Intra-Layer Architectural Characterization.}}
We highlight the following architectural features of kernels for analysis: 1) kernel duration time ($\mathbf{Exe_{Lat}}$); 2) the  read volume $\mathbf{R_v}$ and write volume $\mathbf{W_v}$ through memory bus during kernel execution; 3) input/output data volume ratio ($\mathbf{I_v/O_v}$) where the output volume ($O_v$) equals to the write volume of this kernel and input volume ($I_v$) equals the write volume of the previous executed kernel; 4) kernel dependency distance ($\mathbf{kdd}$), represents the maximum distance in the kernel sequence among current execution kernel and the previous dependent kernels, which can be calculated as follows: $kdd  \approx max(d_{RAW})$.

\begin{figure}[!htbp]
  \centering
  \includegraphics[scale=0.30]{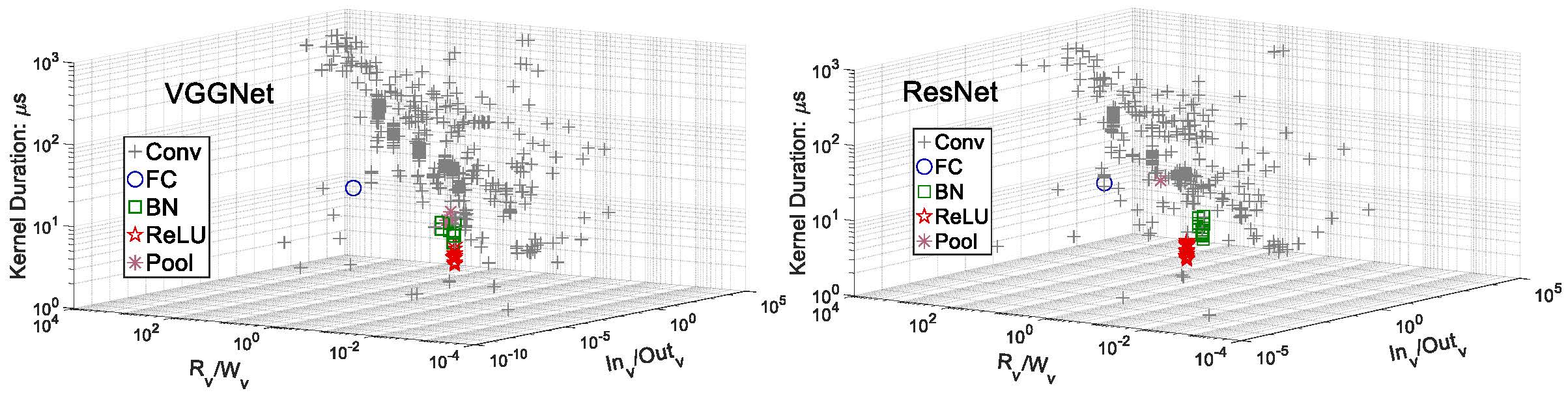}
  \vspace{-20pt}
  \caption{Kernal features of layers.}\label{fig:kernelFeature}
\end{figure}

We observe that although the kernels of different layers have their own features according to their functionality, it is still challenging to predict which layer a kernel belongs to, just based on the these execution features. As shown in  Figure~\ref{fig:kernelFeature}, every point represents the multi-dimensional information ($Exe_{lat}$, $R_v/W_v$, $I_v/O_v$) of an execution kernel. We observe that many points in Figure \ref{fig:kernelFeature} are close to each other which are difficult for identification. Our experiments show that about 30\% of kernels are identified incorrectly with the executed features only and this error rate will increase drastically with more complex network architectures. The detail of the experiments results are shown in Section~\ref{sec:mlp}. 
In summary, the pre-mentioned factors will lower the prediction accuracy for recovering the DNN structure if we only consider the single layer independently. 

\begin{figure*}[!htbp]
  \centering
  \includegraphics[scale=0.35]{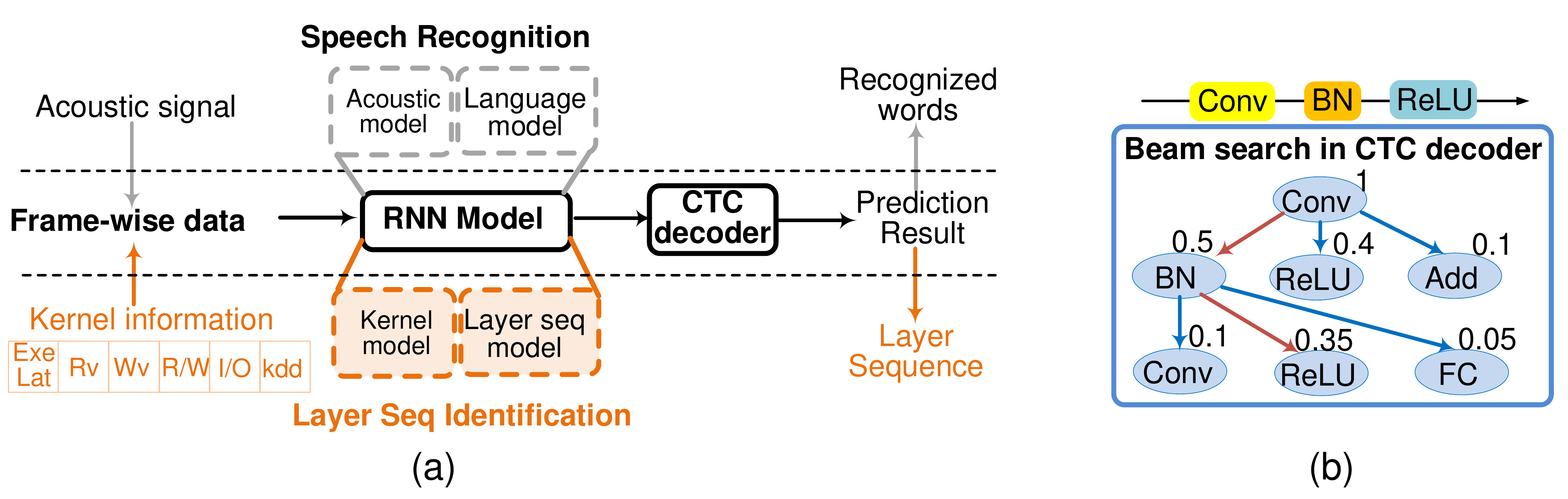}
  \vspace{-5pt}
  \caption{Context-aware layer sequence identification. (a). Map the layer sequence identification to speech recognition problem; (b) An example of layer sequence identification.}\label{fig:LSTM}
\end{figure*}




\noindent\textbf{\ul{DNN Inter-Layer Context.}} Given the previous layer, there is a non-uniform likelihood for the following layer type. This phenomenon provides the opportunity to achieve better layer identification. For example, there is a small likelihood that a FC layer follows a Conv layer in DNN models, because it does not make sense to have two consecutive linear transformation layers.
Such temporal association information between layers (aka. layer context) are inherently brought by the DNN model design philosophy. 
Recalling the design philosophy of some typical NN models, e.g. VGG~\cite{VGG2014}, ResNet~\cite{resnet}, GoogleNet~\cite{GoogleNet}, and DenseNet~\cite{densenet}, there are some common empirical evidences in building network architecture: 1) the architecture consists of several basic blocks iteratively connected. 2) the basic blocks usually include linear operation first (Conv, FC), possibly following normalization to improve the convergence (BN), then non-linear transformation (ReLU), possible down-sampling of the feature map (Pool), and possible tensor reduction or merge (Add, Concat). 

Although DNN architectures evolve rapidly, the basic design philosophy remains the same. Furthermore, the state-of-the-art technical direction of Neural Architecture Search (NAS), which uses reinforcement learning search method to optimize network architecture, also follows the similar empirical experience~\cite{nasnet}. Therefore such layer context generally happens in the network architecture design, which can be used as the prior knowledge for layer identification.

\subsubsection{Run-time Layer Sequence Prediction.} Based on the above analysis,  two major sources of information are jointly considered in layer prediction: the \textbf{\emph{architectural kernel execution features}} and the \textbf{\emph{layer context  distribution possibilities}} in the layer sequence. This problem is similar to the speech recognition, as shown in Figure~\ref{fig:LSTM}, which also involves two parts: the acoustic model converting acoustic signals to text and language models computing text probabilistic distribution in words. Therefore we ingeniously map the run-time layer sequence prediction onto a speech recognition problem and use ASR (auto speech recognition) technologies ~\cite{EndtoEnd_SR, ctc} as a tool to implement the layer identification.  

Formally, the run-time layer sequence prediction problem can be described as follows: We have the kernel execution feature sequence $\vec{X}$ with temporal length of $T$ as an input.  At each time step, kernel feature $\vec{X_t}~(0\leq t<T)$ can be described as a six-dimension tuple: ($Exe_{lat}$, $R_v$, $W_v$, $R_v/W_v$, $I_v/W_v$, kdd)$_t$. The label space $L$ is a set of sequences comprised of all typical layers. The goal is to train an layer sequence identifier $h$ to identify the input kernel feature sequence in a way that minimizes the distance between the predicted layer sequence ($Y$) and oracle sequence ($L$).

\noindent\textbf{\ul{ Context-aware Layer Sequence Identification.}} To build the classifier $h$, we adopt the LSTM model (a typical recurrent neural network) with CTC (Connectionist Temporal Classification) decoder, which is commonly used in Automatic Speech Recognition~\cite{EndtoEnd_SR, ctc}. As shown in Figure \ref{fig:LSTM}, given the input sequence $(\vec{X_1},.., \vec{X_T})$, the output vector $\vec{Y_t}$ is normalized by the softmax operation and transformed to a probability distribution of the next layer OP.  
The object function of training is defined to minimize the CTC cost for a  given target layer sequence $L*$.
\begin{equation}
CTC(X) = -log P(L^*|X)
\end{equation}
where $P(L^*|X)$ denotes the total probability of an emission result $L^*$ in the presence of $X$.

Taking a simplified example in Figure~\ref{fig:LSTM}.(b), there is a sequence within 3 execution kernels. At every time step in ($t_0$, $t_1$, and $t_2$), the LSTM outputs the probability distribution of the layer OPs. At the final time step, the CTC decoder uses beam search to find out the sequence with the highest possibility. In Figure~\ref{fig:LSTM}.(b), the number above an extending node is the total probability of all labelings beginning with this layer OP as the layer sequence prefix. Taking the `BN' at the second row for example, the possibility of sequences with 'BN' as the prefix is 0.5. At every iteration, the extensions of the most probable remaining prefix are explored. Searching ends when a single labeling is more probable than any remaining prefix. In this example, `Conv, BN, ReLU' is the sequence after CTC beam search, which is taken as the prediction result.

At the end of this step, we can get the run-time layer sequence according to the extracted features of the GPU kernel sequence. The experimental details of the model training, validation, and testing are explained in Section \ref{sec:exp}.  

\subsection{Layer Topology Reconstruction}
After obtaining the predicted run-time layer sequence, the next step is to get the connectivity between layers to reconstruct the layer topology. If the feature map data of layer $a$ is fed as the input of layer $b$, there should be a directed topology connection from $a$ to $b$. Since this work focuses on the inference stage, there is only forward propagation across the whole network architecture. 

We first analyze the cache behaviors of feature map data and have the following observations:

\noindent{\textbf{\emph{Observation-1}}}: \ul{Only feature map data (activation data) introduces RAW memory access pattern in the memory bus}. There are several types of data throughout the DNN inference: input images, parameters, and feature map data. Only feature map data is updated during inference. Feature map data will be written to memory hierachy and be read as the input data of the next layer. The input image and parameter data will not be updated during the whole inference procedure. Therefore the RAW memory access pattern will not be introduced by the input image and parameter data, but only by the feature map data.

\begin{figure}[t]
\centering
\includegraphics[scale=0.40]{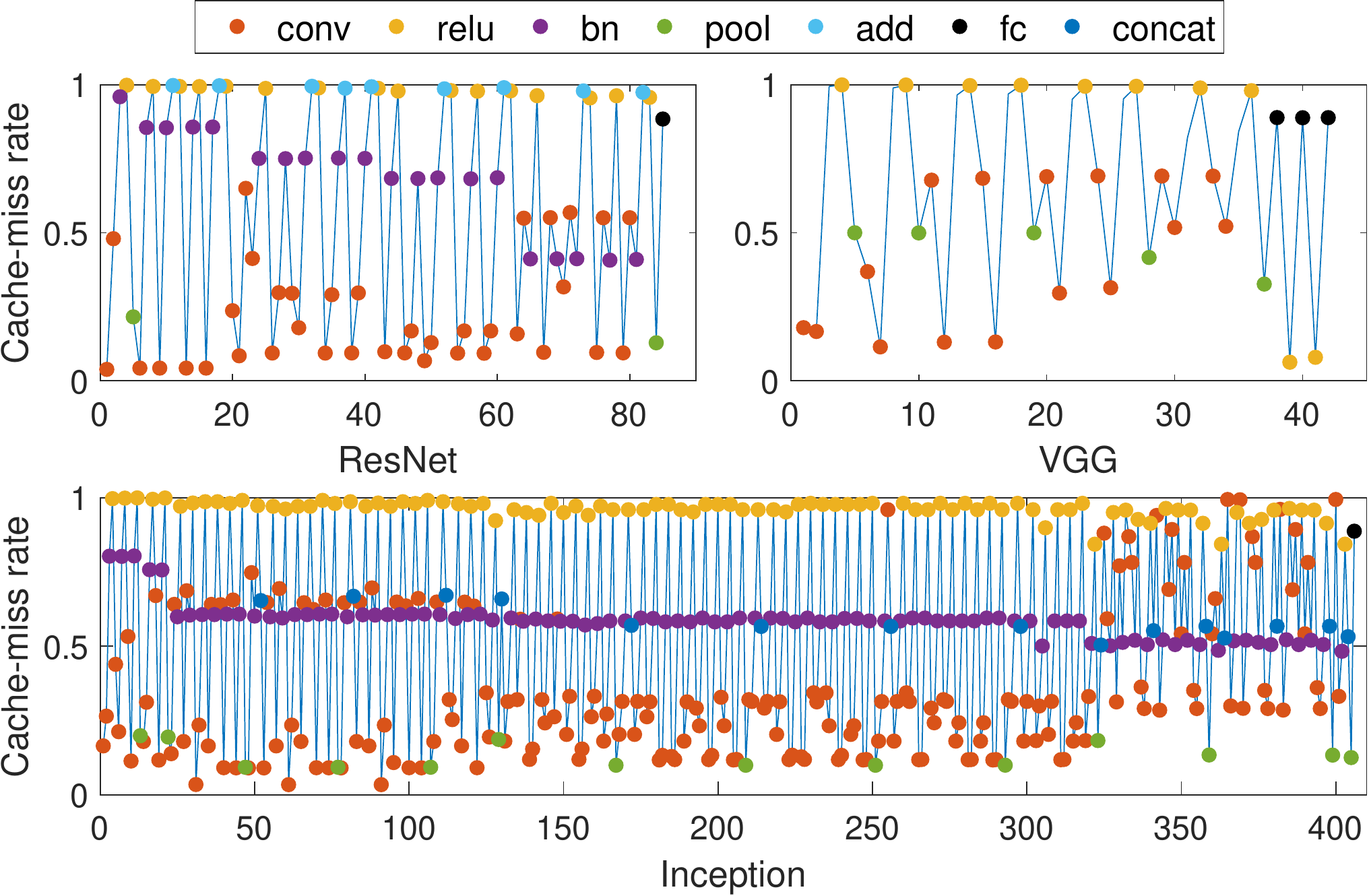}

\caption{Read cache-miss rate of layers in VGG11, ResNet18, and Inception.
}
\label{fig:missrate}
\vspace{-15pt}
\end{figure}

\noindent{\textbf{\emph{Observation-2}}}: \ul{There is very high possibility for the feature map data to introduce read cache misses, especially for the convergent and divergent layers.} 
1) Convergent layer is the one that receives feature map data from several layers. Add and Concat, the main convergent layers, introduce many read cache misses that are contributed by the feature map since they only read the feature map data). As shown in Figure \ref{fig:missrate}, the read cache-miss rate of Add layer is more than 98\% and Concat is more than 50\%. 
2) Divergent layer is the one that output feature map data to several successor layers on different branches. We observe that GPU kernels will execute the layers in one branch before another. Therefore there is a very long distance between this divergent layer and its successor layers in the run-time layer sequence. Because the CUDA library implements extreme data reuse optimization that allocates more cache capacity to the weight tensor instead of the feature map data, it is highly possible that the feature map will be flushed out and need to be read again because of a long reuse distance.  



Based on these two observations, we are able to reconstruct the layer connection by detecting the RAW access patterns in different layers.  We propose a layer topology reconstruction algorithm as follows:
\emph{Step-1}: We scan the memory request address for every layer in the run-time layer sequence. We add the a connection if there is a successor layer reads the same address with the write address of this layer. 
\emph{Step-2}: If there is a non-end layer without any successor, we add the connection between the layer and it next layer in the layer sequence.  

\subsection{Dimension Size Estimation} After the above two steps, we construct the layer topology without the dimension size information. In this section, we explain how to estimate the dimension size parameters according to the memory read and write volume.

\subsubsection{Layer Input/Output Data Volume Estimation} In the first stage, we estimate the input and output size of every layer starting from ReLU layers.

\noindent\textbf{\ul{Step-1: ReLU input/output size estimation.}}
As characterized in the previous subsection, ReLU and Add have high cache miss rate, surpassing 98\%. Hence, the read volume through the bus $R_v$ is almost the same as the input feature map size of the DNN model. Then the write volume $W_v$ can be estimated which is equal to $R_v$. Based on this observation, we can obtain the input and output size of ReLU layers. 

\noindent\textbf{\ul{Step-2: Broadcasting ReLU size to other layers.}}
In neural network, the previous layer's output acts as the input to current layer, so the output size  (feature map height/width and channel number for Conv or neuron number for FC) of the previous layer equals to the input size of current layer. Hence, given the input size of a ReLU layer, the output size of the previous BN/Add/Conv/FC layer and the input size of the next Conv/FC layer can be estimated. Since the ReLU layer is almost a standard layer every basic blocks,
it can guide the dimension size estimation of its adjacent layers. The Add layer can play a similar role for dimension estimation at the divergence and convergence points of compute branches.

\noindent\textbf{\ul{Step-3: Estimate the DNN input and output size.}} In this step, we estimate the input size of the first layer and output size of the last layer with the PCIe information. As described in Section~\ref{sec:pcie_data}, the adversary is able to get the memory copy size through PCIe. The input image data is copied to GPU at the beginning and prediction results data is copied to host at the end of a batch inference. Therefore the input size and output size can be inferred from $Mem_{cp}$.

\subsubsection{Dimension Space Calculation}
In the second stage, we calculate the dimension parameters with the constructed the layer topology knowing the input/output size of every layer ( $I_{i}/O_{i}$). 
We want to estimate the following dimension space: the input (output) channel size $IC_{i}$ ($OC_i$), the input (output) height $IH_i$ ($OH_i$), the input (output) width $IW_i$ ($OW_i$), the the weight size ($K\times K$), and the convolution padding $P$ and stride $S$. In fully connected layers, $OC$ denotes the neuron number. The quantitative estimation is listed in Table~\ref{tab:dim}. Based on the fact that the input size of each layer keep the same as the output size of previous layer, and following the constraints shown in Table~\ref{tab:dim}, we are able to search the possible solutions for every layer. 

\begin{table}[h]
 \caption{Dimension space calculation.}\label{tab:dim}
\centering
 \scalebox{0.95}{
\renewcommand\arraystretch{0.8}
    \begin{tabular}{|c|c|c|}
    \hline
    \normalsize \textbf{Layer OP} & \normalsize \textbf{Constraints} \& \textbf{Estimation} \\ \hline
    \normalsize \multirow{3}*{\textbf{Conv}} & $OH_i = \lfloor(IH_i+2P-K)/S\rfloor+1$   \\
    & $OW_i = \lfloor(IW_i+2P-K)/S\rfloor+1$ \\
    & $OH_i\times OW_i\times OC_i = O_i/N$ \\ \hline
    \normalsize \multirow{3}*{\textbf{Pool}} & $OH_i = \lfloor(IH_i+2P-K)/S\rfloor+1$   \\
     & $OW_i = \lfloor(IW_i+2P-K)/S\rfloor+1$ \\
     & $OC_i=IC_i$, $OH_i\times OW_i\times OC_i = O_i/N$ \\ \hline
    \normalsize \textbf{FC} &  $OC_i = Oi/N$ \\ \hline 
    \normalsize \textbf{BN}	&  $OH_I$ = $IH_i$,
$OW_I$ = $IW_i$,
$OC_i$ = $IC_i$ \\ \hline
   \normalsize \textbf{ReLU} & $OH_I$ = $IH_i$,
$OW_I$ = $IW_i$,
$OC_i$ = $IC_i$ \\ \hline
   \normalsize \textbf{Add} &   $OH_i$ = $IH_{i_j}$,
$OW_i$ = $IW_{i_j}$,
$OC_i$ = $IC_{i_j}$ \\ \hline
   \normalsize \textbf{Concat} &  $OH_i$ = $IH_{i_j}$, $OW_i$ = $IW_{i_j}$, $OC_i=\sum_{j}IC_{i_j}$\\ \hline   
    \end{tabular}
    }
\end{table}

\section{Futher-step Adversarial Attack}~\label{sec:adv_flow}

The extracted network architecture can be used to conduct further-step attack. In this work, we use the adversarial attack as a use case to show the importance of network architecture, which is also one of the most common attack means in the domain of neural network security.

In the adversarial attack, the adversaries manipulate the output of the neural network model by inserting small perturbations to the input images that still remain almost imperceptible to human vision~\cite{fgsm}. 
 The  goal of adversarial attack is to search the minimum perturbation on input that can mislead the model to produce an arbitrary (\emph{untargeted attack}) \cite{fgsm} or a pre-assigned   (\emph{targeted attack}) \cite{intriguing:szegedy2013,targetRegressionAttack:AAAI,physicalMLattack:2016} incorrect output. 
To conduct the adversarial attack against a black-box model,  the adversary normally builds a substitute model first, by querying the input and output of the victim model. Then the adversary generates the adversarial examples based on the white-box substitute model~\cite{intriguing:szegedy2013,Papernot:2017:blackbox,axiv:papernot:transferability}. Finally, they use these adversarial examples to attack the black-box model. 

\begin{figure}[t]
\centering
\includegraphics[scale=0.50]{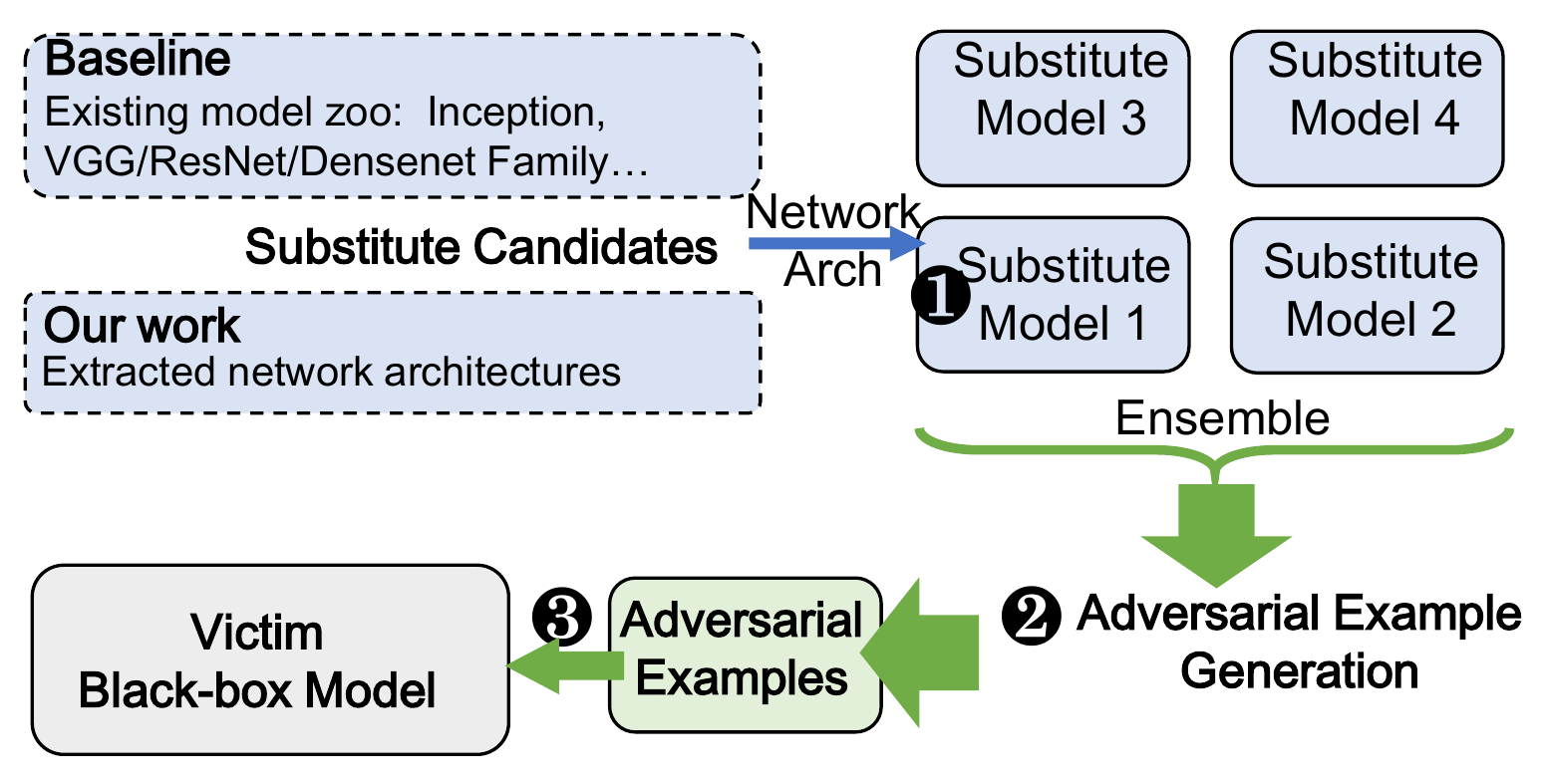}

\caption{Adversarial Attack Flow.
}
\label{fig:adv_flow}
\vspace{-15pt}
\end{figure}

\noindent\textbf{\ul{Step-1: Build substitute models}}
In our work, we train substitute models with the extracted network architectures,
while previous work select the typical network architectures to build the substitute model, as shown in Figure~\ref{fig:adv_flow} .

\noindent\textbf{\ul{Step-2: Generate adversarial examples}}
The state-of-art solution \cite{ICLR17:Delving:Dawn} uses ensembled method to improve the attacking successful rate based on the hypothesis that if an adversarial image remains adversarial for multiple models, it is more
likely to be effective against the black-box model as well. We follow the similar techniques to generate adversarial images for the ensemble of multiple models. 


\noindent\textbf{\ul{Step-3: Apply the adversarial examples}} As the final step, the adversary attacks the black-box model using the generated adversarial examples as input. 

We follow the same adversarial attack methodology in the previous work~\cite{ICLR17:Delving:Dawn} and the only difference is that we use the predicted network architecture to build the substitute models and the experiments show that with the accurate extracted network architecture, the successful rate of adversarial attack will be improved significantly. The detailed results will be shown in Section~\ref{sec:adv_result}.

\section{Attack Demonstration}\label{sec:exp}   
In this section, we evaluate a complete hacking flow, taking CNNs as a case study. 

\subsection{Evaluation Methodology}\label{sec:setup}
\noindent\textbf{\ul{Experimental Platform:}} To validate the feasibility of stealing the memory information during inference execution, we conduct the experiments on the hardware platform equipped with Nvidia K40 GPU \cite{NVIDIA_K40}. The NN models are implemented based on Pytorch framework \cite{pytorch}, with CUDA8.0 \cite{cuda} and cuDNN optimization library \cite{cuDNN}. 

\noindent\textbf{\ul{Experimental Setup:}}
We use the nvprof tool to emulate bus snooping. nvprof is an NVIDIA profiling tool kit that enables us to understand and optimize the performance of OpenACC applications \cite{nvprof}. The information that we can get from the nvprof tool is shown in Table \ref{tab:info_summary}. Based on these raw data, we can reconstruct the network architecture of the victim model according to the steps in Figure \ref{fig:systemstack}.(b).

\noindent\textbf{\ul{Layer Sequence Identifier Training:}} As an initial step for network architecture extraction, we first train the layer sequence identifier based on an LSTM-CTC model for layer  sequence identification. The detailed training procedure is as follows.

\emph{Dataset for Training:} In order to prepare the training data, we first generate 8500 random computational graphs and obtain the kernel features experimentally with nvprof which emulates the process of bus snooping.  Two kinds of randomness are considered during random graph generation: topological randomness and dimensional randomness. At every step, the generator randomly selects one type of block from sequential, Add, and Concat blocks. The sequential block candidates include (Conv, ReLU), (FC, ReLU), and (Conv, ReLU, Pool) with or without BN. The FC layer only occurs when the feature map size is smaller than a threshold. The Add block is randomly built based on the sequential blocks with shortcut connection. The Concat block is built with randomly generated subtrack number, possibly within Add blocks and sequential blocks. The dimensional size parameters, such as the channel, stride, padding, and weight size of Conv and neuron size of FC layer,  are randomly generated to improve the diversity of the random graphs. The input size of the first layer and the output size of the last layer are fixed during random graph generation, considering that they are usually fixed in one specific target platform. 
We randomly select 80\% of the random graphs as the \textbf{\emph{training set}} and other 20\% as the  \textbf{\emph{validation set}} to validate whether the training is overfitting or not. To verify the effectiveness and generalization of our hardware-aided framework, we examine various NN models as the \textbf{\emph{test set}}, including  VGG \cite{VGG2014}, ResNet \cite{resnet}, Inception \cite{GoogleNet}, and Nasnet~\cite{nasnet} to cover the layer types as many as possible.

\emph{Identifier Configurations:} The identifier utilizes the LSTM-CTC model, consisting of one hidden layer with 128 cells, for layer sequence identification. It is trained using the Adam \cite{adam} optimizer with learning rate adaptation. The training is terminated after 100 epochs. 

\subsection{Run-Time Layer Sequence Identification}
In this section, we first evaluate the layer sequence identification accuracy. Then we analyze the importance of the layer context information and how does the noise affect the identification accuracy. 

\subsubsection{Prediction Accuracy} 

\underline{\emph{Evaluation Metric.}}
The speech recognition model adopts the mean normalized edit distance between the predicted sequence and label sequence to quantify the prediction accuracy \cite{ctc, EndtoEnd_SR}, which is referred to as label error rate (LER). Therefore, here we also adopt LER to evaluate the prediction accuracy. The detailed LER calculation is formulated as the following equation ~\cite{ctc}.

\begin{equation}
LER = \frac{1}{|S|} \sum_{(x,z)\in S} \frac{ED(h(x),z)}{|z|}
\end{equation}
where $ED(p,q)$ is the edit distance between two sequences $p$ and $q$, i.e. the minimum number of insertions,
substitutions, and deletions required to change p
into q, $|S|$ is the number of samples in testing set, $h(x)$ is the identified layer sequence, and $z$ is the oracle layer sequence. 

\underline{\emph{Results.}} We first evaluate the accuracy on validation set. 
The average LER on validation set is about 0.08, which evidences the good prediction capability. Furthermore, we evaluate the accuracy to identify several typical networks, as shown in Table \ref{tab:summary}.   For VGG and ResNet families, the prediction LER is lower than 0.07. For inception and Nasnet, the LER increases a little bit because of the much deeper and complex topology. In summary, our proposed method predicts generally well in these cases.

\underline{\emph{A Detailed Example.}}
We take ResNet34 as an example to present the detailed results in Table \ref{tab:test_example}. We make the following observations: 1) The prediction model is generally effective in correctly identifying the layer sequence; 2) In some rare cases, although it is possible that the BN/ReLU will be incorrectly missed or created, the critical Conv/FC/Add/Concat layers can be correctly recognized.

\begin{table}[h]
    \caption{Prediction LER on typical networks.}\label{tab:summary}
\centering
\renewcommand\arraystretch{1.0}
\scalebox{0.65}{
    \begin{tabular}{|c|c|c|c|c|c|c|}
    \hline
    VGG16 & VGG19 & ResNet34 & ResNet101 & ResNet152& Inception & Nasnet\_large \\ \hline
    0.020 & 0.017 & 0.040 & 0.067 & 0.068 & 0.117 & 0.132 \\ \hline
    \end{tabular}
    }
\end{table}

\begin{table}[h]
 \caption{Detailed comparison between the oracle and predicted layer sequence.}\label{tab:test_example}
\centering
\renewcommand\arraystretch{0.8}
 \scalebox{0.7}{
    \begin{tabular}{| p{1.5cm} | p{4.2cm} | p{4.2cm} |}
    \hline
    \normalsize \textbf{Network} & \normalsize \textbf{Oracle Sequence} &  \normalsize \textbf{Predicted Sequence}  \\ \hline
  
    
    ResNet34 (LER 0.040) & conv bn relu pool conv bn relu conv bn add relu conv bn relu 
conv bn add relu conv bn relu conv bn add relu conv bn relu  conv bn conv bn add 
relu conv bn relu conv bn add relu conv bn relu conv bn add relu conv bn relu 
conv bn add relu conv bn relu conv bn conv bn add relu conv bn relu conv bn 
add relu conv bn relu conv bn add relu conv bn relu conv bn add relu conv bn
relu conv bn add relu conv bn relu conv bn add relu conv bn relu conv bn 
 conv bn add relu conv bn relu conv bn add relu conv bn relu conv bn add relu pool fc & conv {\color{red}{\sout{bn}}} relu pool conv bn relu conv bn add relu conv bn relu conv bn add relu conv bn relu conv bn add relu conv bn relu conv bn conv bn add relu conv bn relu conv bn add relu conv bn relu conv bn add relu conv bn relu conv bn add relu conv bn relu conv bn conv bn add relu conv bn relu conv bn add relu conv bn relu conv bn add relu conv bn relu conv bn add relu conv bn relu conv bn add relu conv bn relu conv bn add relu conv bn relu conv bn conv bn add relu conv bn relu conv bn {\color{red}relu} add relu conv {\color{red}{\sout{bn}}} relu conv bn {\color{red}relu} add relu pool fc \\ \hline
    \end{tabular}
    }   
\end{table}

\subsubsection{With/Without Layer Context Information.}\label{sec:mlp} 

\begin{figure}[!htbp]
  \centering
  \includegraphics[scale=0.25]{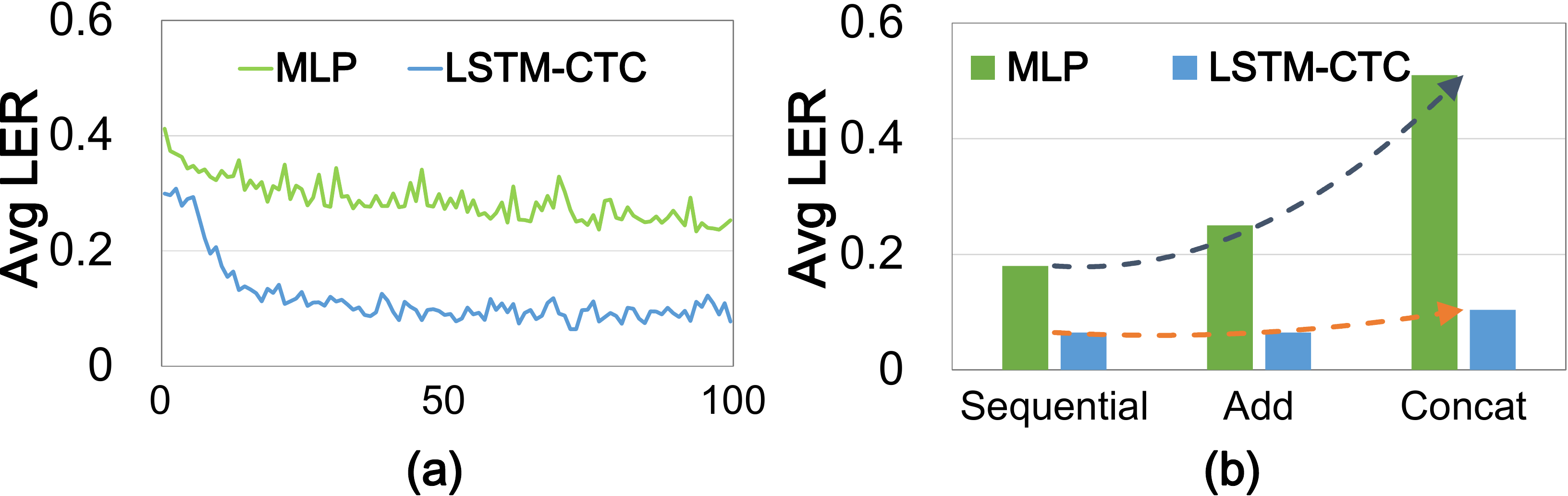}

  \caption{MLP: layer prediction without inter-layer information; LSTM-CTC: leverage the inter-layer information for better prediction. The results indicate that the inter-layer likelihood is very important for layer prediction.}\label{fig:predict_ana}
\end{figure}

We analyze the importance of inter-layer information in this section. As shown in Figure \ref{fig:predict_ana}(a), we compare the LER of two methods: layer-context-aware identifier based on LSTM-CTC model and single-layer identifier based on MLP (multi-layered perceptron) model. 

We draw two conclusions from this experiment: 1) We can achieve much better prediction accuracy with considering the layer context information. The results show that the average LER of LSTM-CTC is two times lower than the MLP-based method. 2) Layer context information is increasingly important when identifying more complex network architecture.   As shown in Figure \ref{fig:predict_ana}.(b), compared to the simple network architecture with only chain typologies, the more complex architectures with remote connections (e.g. Add or Concat) cause higher error rates. For the MLP-based model, the LER dramatically increases when the network is more complex (from 0.18 to 0.5); while for the LSTM-CTC model, the average LER demonstrates a non-significant increase (from 0.065 to 0.104).

The experiments results indicate that the layer context with inter-layer temporal association is a very important information source, especially for the layer sequence within complex topology. 

\subsubsection{Noise Influence.}\label{sec:robustness}

We conduct experiments to analyze the prediction sensitivity in the scenarios with noise on the kernel features. When 5\%, 10\%, 20\%, or 30\% of random noises is inserted to the read and write volumes of the validation set, the average LERs of the layer prediction is shown in Table~\ref{tab:noise}. The results indicate that the layer sequence identifier has the ability to resist noise.

\begin{table}[h]
    \caption{Prediction LER on validation set with random noise.}\label{tab:noise}
\centering
\renewcommand\arraystretch{1.0}
\scalebox{0.9}{
    \begin{tabular}{|c|c|c|c|c|c|c|}
    \hline
     Noise & 5\% & 10\% & 15\% & 30\% \\ \hline
    LER & 0.08 & 0.09 & 0.12 & 0.16\\ \hline
    \end{tabular}
    }
\end{table}


\underline{}

 \begin{figure}[!htbp]
  \centering
  \includegraphics[scale=0.55]{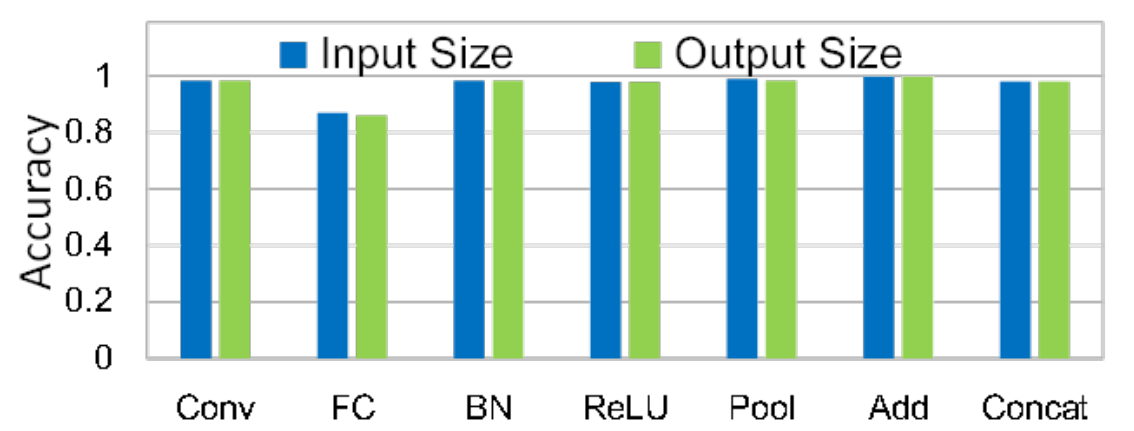}
  \caption{Layer input and output size estimation (normalized to the oracle size).}\label{fig:iosize}

\end{figure}

\begin{figure*}[!htbp]
  \centering
  \includegraphics[scale=0.42]{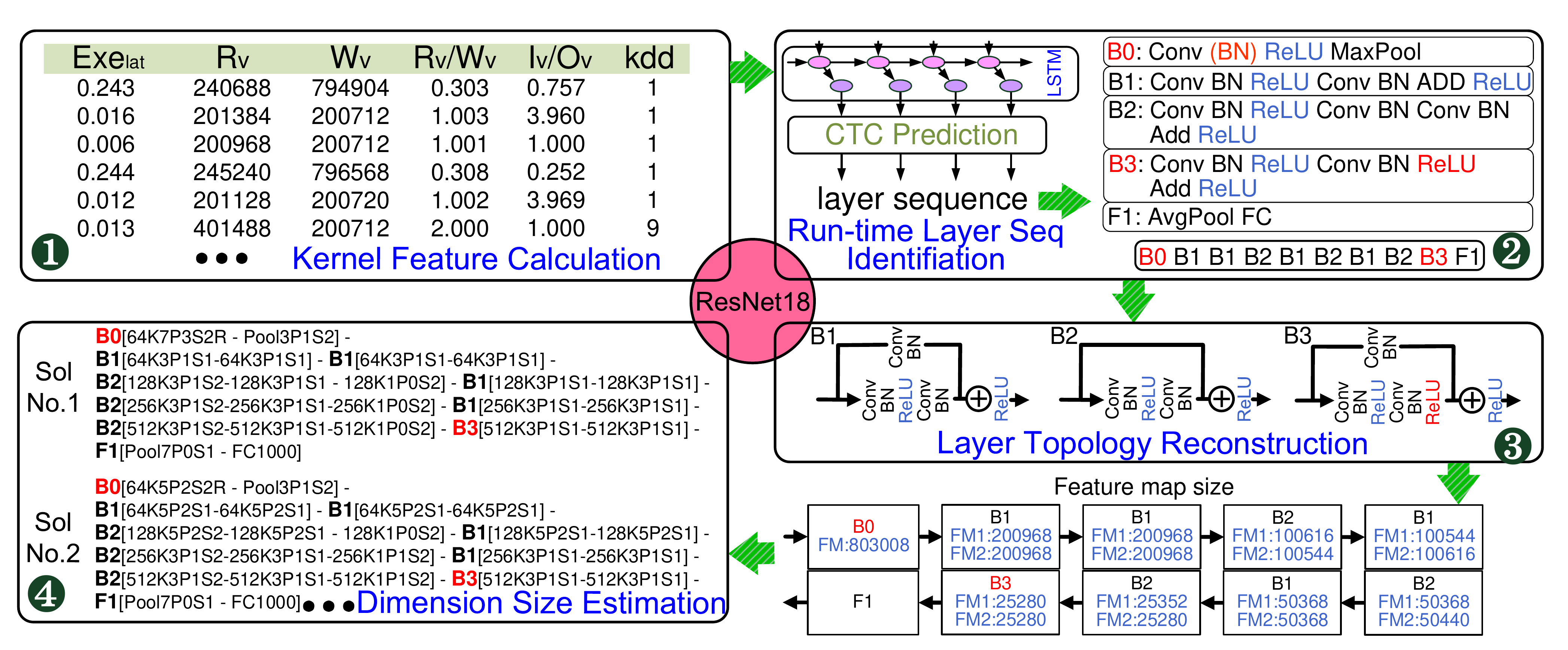}
  \caption{Illustration of a complete example of reconstructing ResNet18. Step-1: Calculating the kernel features based on the bus snooped information (emulated by nvprof); Step-2: Identify the run-time layer sequence; Step-3: Reconstruct the layer topology and get the input and output size of layers; Step-4: dimension estimation according to the layer type and input/output size.}\label{fig:example}
\end{figure*}

\subsection{Dimension Space Estimation}
In this section, we show how accurate the size estimation of the input and output for every layer can be, which is important for dimension space estimation. We take the input and output size of every layer as examples and show the results in Figure \ref{fig:iosize}. The estimated size is normalized to the oracle size. For Conv, BN, ReLU, Add, and Concat, the estimation accuracy can reach up to 98\%. The FC presents lower accuracy, because of the following reasons: the FC layer is usually at the end of the network and the neuron number shrinks. Therefore, the activation data of the ReLU layer will probably be  filtered, and the corresponding cache miss rate will be much lower. Therefore, it is not  accurate to use ReLU read transactions to estimate the FC size. Instead, we use the read volume of FC layer to predict the input and output size.

\subsection{An end-to-end Demo Case}
We use a complete example to clearly illustrate the model extraction against the victic model ResNet18. With the extracted network architecture, we conduct the consequent adversarial attack which shows that the attack success rate can be significantly improved. 

\subsubsection{Model Extraction}
The model extraction flow is shown in Figure \ref{fig:example}, consisting the following four steps. 

\noindent\textbf{\ul{Step-1: Kernel feature calculation:}} 1) We obtain the bus snooped information and calculate the sequence of kernel features ($Exe_{Lat}$, $R_v$, $W_v$, $R_v/W_v, In/Out$), as shown in Subgraph-{\small\circled{1}}.

\noindent\textbf{\ul{Step-2: Run-time layer sequence identification:}} Taking in the kernel feature sequence, we use the layer sequence identifier, which is based on an LSTM-CTC model, to identify the layer sequence. The prediction results are listed in the right boxes of Subgraph-{\small\circled{2}}. 
Because the network architecture usually constitutes of several basic blocks iteratively, we use a hierarchical expression to show the prediction results to facilitate the presentation. The complete layer sequence of ResNet18 consists of \{$B_0, B_1, B_1, B_2, B_1, B_2, B_1, B_2, B_3, F_1 $\}. The identifier predicts the layer sequence precisely for most of the blocks and there are only small mistakes in two blocks: In $B_0$, the identifier missed a BN layer; In $B_3$, the identifier incorrectly adds another ReLU in the architecture. Since there is no change to the feature map size in BN and ReLU layers, this prediction will not affect the dimension size estimation results.

\noindent\textbf{\ul{Step-3: Layer topology reconstruction:}} Based on the RAW memory access dependency pattern, the layer topology is constructed following the reconstruction algorithm, as shown in Subgraph-{\small\circled{3}}.

\noindent\textbf{\ul{Step-4: Dimension size estimation:}} 1) We first estimate the feature map input and output size of ReLU layers according to their read and write volume (in blue color); 2) 
We then estimate the input and output size of all the other layers propagating from ReLU; 
3) We finally estimate the complete dimension space based on the input and output size of layers, according to the equations in Table~\ref{tab:dim}. With reasonably assuming the kernel and stride size, the padding, input channel, output channel in each Conv Layer can be consequently derived.

Two examples of the possible dimension solutions are shown in  Subgraph-{\small\circled{4}}. 
`[]' represents a basic block with an Add layer; `K', `P', and `S' represent the weight, the padding, and the stride size of the Conv layer, respectively; `B' represents BN; and `R' represents ReLU. 
 
 We successfully reconstructed the network architecture of the black-box DNN model after these four steps. Although the final graph is not unique due to the variable dimension size, they are within the same network architecture family. We will show the importance of these predicted network architectures which can help boost the adversarial attacking efficiency in the following subsection.

\subsubsection{Adversarial Attack Efficiency}\label{sec:adv_result}
In this section, we show that the adversarial attack efficiency can be significantly improved with the extracted network architecture information. 

The adversarial attacking flow consists of three steps: 1) building substitute models; 2) generating adversarial examples towards the substitute model; 3) applying the adversarial examples to the victim black-box model, as introduced in Section~\ref{sec:adv_flow}. We use the algorithm proposed in prior work~\cite{ICLR17:Delving:Dawn}, which achieves better attacking success rate by generating the adversarial examples based on the ensemble of four substitute models. 

\noindent\textbf{\ul{Setup:}}
In these experiments, we use ResNet18~\cite{resnet} as the victim model for targeted attack.
Our work adopts the extracted neural network architecture, as shown in Figure~\ref{fig:example}, to build the substitute models. For comparison, the baseline examines the substitute models established from following networks:
\texttt{VGG} family (\texttt{VGG11}, \texttt{VGG13}, \texttt{VGG16}, \texttt{VGG19})~\cite{VGG2014},
 \texttt{ResNet} family (\texttt{ResNet34}, \texttt{ResNet50}, \texttt{ResNet101}, \texttt{ResNet152})~\cite{resnet}, 
\texttt{DenseNet} family (\texttt{DenseNet121}, \texttt{DenseNet161}, \texttt{DenseNet169}, \texttt{DenseNet201})~\cite{densenet},
\texttt{SqueezeNet}~\cite{squeezenet}, and \texttt{Inception}~\cite{GoogleNet}.

\noindent\textbf{\ul{Results:}} First,  we randomly select 10 classes, each class with 100 images from ImageNet dataset~\cite{deng2009imagenet} for testing. To perform the targeted attack, we test both the cases that the targeted class is far away from the original class and the targeted class is close to original class. We compare the following five solutions: ensembled model with substitute models from VGG family, DenseNet family, mix architectures (squeezeNet, inception, AlexNet, DenseNet), ResNet family, and from extracted ResNet architectures using our previous model extraction. The results are shown in Table~\ref{tab:transferability}. We get several observations: 1) The attacking success rate is generally low for the cases without network architecture knowledge. The adversarial examples generated by substitute model with VGG family, DenseNet family, and mix architectures only conduct successful attacks in 14\%--25.5\% of the cases. 2) With some knowledge of the victim architecture, the attacking success rate can be significantly improved. For example, the substitute models within ResNet family can achieve attack success rate of 43\%. 3) With accurate network extraction, although it still has a little difference from the original network, the attacking efficiency can be boosted to 75.9\%. These results indicate that our model extraction can significantly improve the success rate of consequent adversarial attacks. 

\begin{figure}[!htbp]
  \centering
  \includegraphics[scale=0.35]{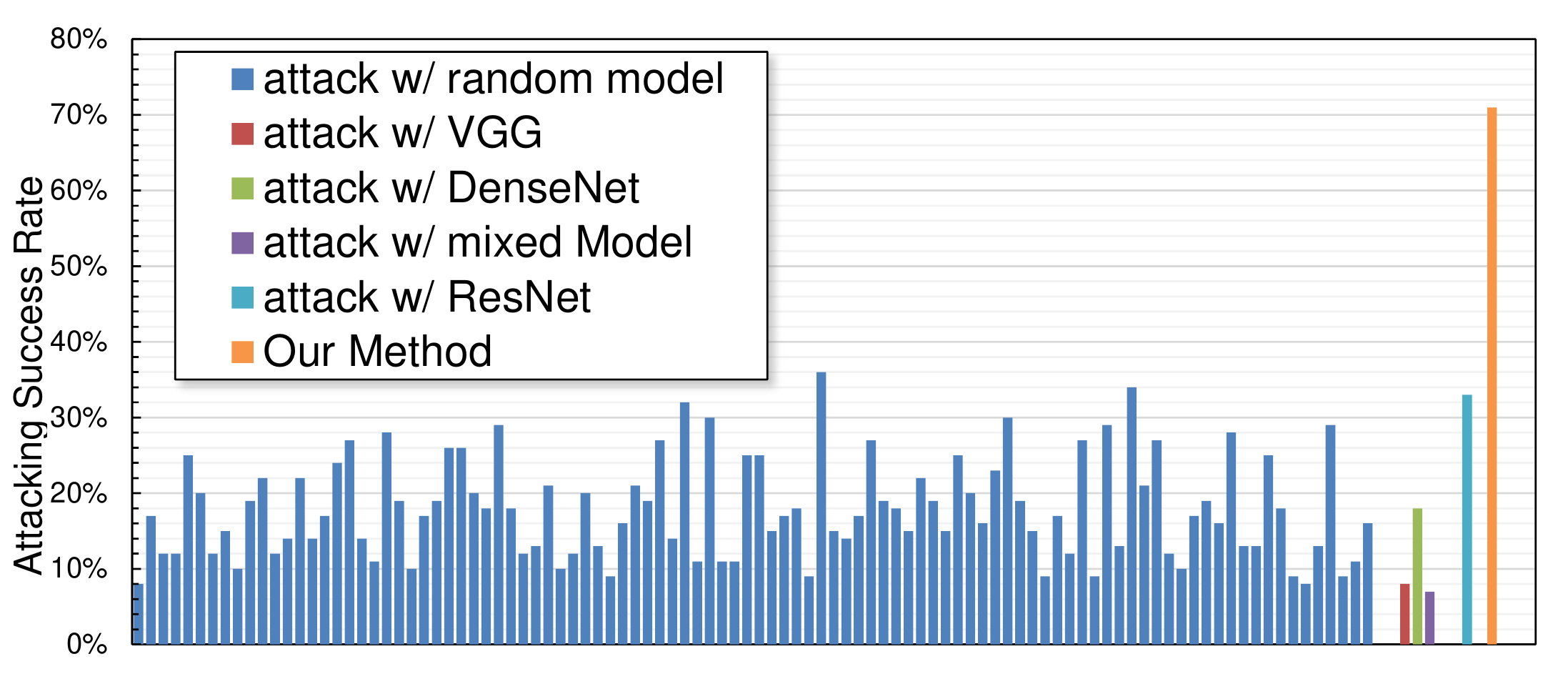}
  \caption{Explore attack success rate across different cases when conduct targeted attack (Class-750 $\rightarrow$ Class-255): 1) random picking 4 subsitute modes from candidate pool; 2) substitute models from VGG family; 3)  substitute models from DenseNet family; 4)  substitute models from squeezeNet, inception, AlexNet, and DenseNet (Mix); 5)  substitute models from ResNet family ; 6)  substitute models from our extracted network architectures.}\label{fig:compare_trans}
\vspace{-5pt}
\end{figure}

 In a further step, we take a deep look at the targeted attack leading the images in Class-755 to be misclassified as Class-255, in order to explore the ensembled model with various substitute combinations. We randomly pick four substitute models from the candidate model zoo and the results are shown in the blue bars of Figure~\ref{fig:compare_trans}. We also compare the results to the cases using substitute models 1) from VGG family; 2) from DenseNet family; 3) from squeezeNet, inception, AlexNet, and DenseNet ('Mix' bar in the figure); 4) from ResNet family; and 4) from extracted cognate ResNet18 model (Our method)  to generate the adversarial examples. As shown in Figure~\ref{fig:compare_trans}, the average success rate of random cases is only 17\% and the best random-picking case just achieves the attacking success rate of 34\%. We observe that all good cases in random-picking (attacking success rate > 20\%) include substitute models from ResNet family. Our method with accurate extracted DNN models performs best attacking success rate across all the cases, 40\% larger than the best random-picking case and ResNet family cases. In a nutshell, with the help of the effective and accurate model extraction, the consequent adversarial attack can achieve much better attack success rate.
 

   \begin{table}[h]
    \caption{Success rate with different substitute models.}\label{tab:transferability}
\centering
\renewcommand\arraystretch{1.0}
\scalebox{0.8}{
    \begin{tabular}{|c|c|c|c|c|c|c|}
    \hline
    & VGG  & DenseNet  & Mix & ResNet & Extracted \\ 
    & family & family & & family & DNN \\ \hline
    Success rate & 18.1\% & 25.5\% & 14.6\% & 43\% & 75.9\%  \\ \hline
    \end{tabular}
    }
\end{table}

\section{Discussion}\label{sec:discussion} 

In this section, we first analyze the intrinsic reason of successful network architecture extraction and how general it will work. Then, we further explain the impact of the existing memory bus protection methods on our attacking approach and propose potential defensive techniques.


\subsection{Generality and Insights}

The standardization through the whole stack of neural network system facilitates the model extraction.  The standardized hardware platforms, drivers, libraries, and frameworks are developed to help machine learning industrialization with user-friendly interfaces. Transforming from the input neural network architecture to final hardware code is dependent on the compilation and scheduling strategies of DNN system stacks, which can be learned under the same execution environment.  Therefore, the adoption of these hardwares, frameworks, and libraries in the development workflow gives attackers an opportunity  to investigate the execution pattern and reconstruct the network architecture based on the hardware execution information. 

The root cause of hacking the network structure is to learn the transformations between framework-level computation graphs and kernel feature sequence. Therefore, we build the training set based on random graphs with basic operations provided by DNN framework. In our methodology, as long as the DNN model is built based on the basic operations provide by framework (such as Conv2d, ReLU, and MaxPool2d, etc in pytorch), the neural network structure can be reconstructed.  In addition, our methodology can be extended to include the other operations in the framework model zoo. Therefore, our methodology is generally applicable to various CNN models with different neural network architectures. We demonstrate that memory address traces are able to damage the NN system security which urges hardware security studies(e.g.ORAM) and may raise the attention of the architecture/system community to build more robust NN system stack.



\subsection{Defence Strategies} 
\subsubsection{Microarchitecture Methodologies.}
There are a few architectural memory protection methods. 
\underline{\emph{Oblivious Memory:}} To reduce the information leakage on the bus, previous work proposes oblivious RAM (ORAM) \cite{path_oram:2013,Liu:ORAM, Liu:2013:ORAM}, which hides the memory access pattern by encrypting the data addresses. With ORAM, attackers cannot identify two operations even when they are accessing the same physical address~ \cite{path_oram:2013}.  However, ORAM techniques incur a significant memory bandwidth overhead (up to an astonishing 10x), which is impractical to be used on GPU architecture that is  bandwidth sensitive.

\underline{\emph{Dummy Read/Write Operations:}} Another possible defence solution  is to introduce  fake memory traffic to disturb the statistics of memory events.  Unfortunately the noise exerts only a small degradation of the layer sequence prediction accuracy, as illustrated in Section~\ref{sec:robustness}. As such, dummy RAW operations to obfuscate the layer dependencies identification may be a more fruitful defensive technique to explore. 


\subsubsection{System Methodologies.}
The essence of our work is to learn the compilation and scheduling graphs of the system stack.  Although the computational graphs go through multiple levels of the system stack, we demonstrate that it is still possible to recover the original computational graph based on the raw information stolen from the hardware.  At the system level one could:  1) customize the overall NN system stack with TVM,  which is able to implement the graph level optimization for the operations and the data layout~\cite{tvm}. The internal optimization possibly increases the difficulty for the attackers to learn the scheduling and compilation graph, or 2) make secure-oriented scheduling between different batches during the front-end graph optimization. Although such optimizations may have little impact on performance, they may obfuscate the attackers view of kernel information.

\section{Related Work}\label{sec:related}
Machine learning security is an attractive topic with the industrialization of DNNs techniques. The related existing work mainly comes from the following two aspects.

\noindent\textbf{\ul{Algorithm perspective:}} Neural network security attracts much attention with the industrialization of the DNN techniques. Previous work discuss the concrete AI security problems and machine learning attack approaches~\cite{ucb, NNsafetysummray:papernot2016, mitAttackSummary:18}. Adversarial attacking are one of the most important attack model which generates the adversarial examples with invisible perturbation to confuse the victim model for wrong decision. These adversarial examples can produce either the targeted \cite{intriguing:szegedy2013, fgsm, CoRR15:Papernot, carlini2017towards, baluja2017adversarial, das2011differential, cisse2017houdini, sarkar2017upset}, or untargeted \cite{physicalMLattack:2016, deepfool:cpvr2016, moosavi2017universal, su2017one, mopuri2017fast} output for further malicious actions.  The adversarial attacks can be categorized as {white-box attacks} \cite{KDD2013:evasion,  intriguing:szegedy2013, CoRR15:Papernot, realFaceDetactionAttack:2016, physicalMLattack:2016,malwareClassification:2016,targetRegressionAttack:AAAI,fgsm, ateniese2013hacking, sabour2015adversarial, rozsa2016adversarial} and {black-box  attacks} \cite{stealing_para, axiv:papernot:transferability, fgsm, physicalMLattack:2016, intriguing:szegedy2013, Papernot:2017:blackbox, PracticalEvasion, shokri2017membership, xu2016automatically, fredrikson2014privacy, narodytska2016simple, joon2017adversarial, mopuri2017fast, dong2017boosting, carlini2017ground, cisse2017houdini, sarkar2017upset}, according to the prior knowledge regarding the victim model. More specifically, in white-box attack, attacker knows internal model characteristics (i.e. network architecture and parameter of the model) of the victim model. White-box attack is less practical than black-box attack in real deployment since the designers intend to hind the information from the users. In black-box attack, the attacker has no knowledge of the model characteristics but can only query the black-box model for the input and output responses.
The state-of-art work observes that adversarial examples transfer better if the substitute and victim model are in the same network architecture family~\cite{ICLR17:Delving:Dawn,meta_learning}. Therefore, the extracting inner network structure is important for attacking effectiveness.

Consequently, model extraction work are emerged to explore the model characteristics. Previous work steal the parameter and hyperparameter of DNN models with the basic knowledge of NN architecture~\cite{stealing_para,stealing_hyperpara}.  \emph{Seong et al.}  explore the internal information of the victim model based on meta-learning~\cite{meta_learning}. However, it is inefficient to extract the neural network architecture in the algorithm level, Our work addresses this issue from system perspective with enhancement of bus snooped  information, which will significantly benefit the attack efficiency of current software algorithms.






\noindent\textbf{\ul{Hardware perspective:}} Several accelerator based attacks are proposed, either aiming to conduct model extraction ~\cite{DAC18:ReverseNN} or input inversion~\cite{ICCAD:Qiang}. However, their methodology relies on the specific features in hardware platforms and cannot be generally applicable to GPU platforms with full system stack.  
Some studies research on the information leakage in general purpose platforms. Cathy~\cite{arXiv18:cacheTelpathy} explores side-channel techniques to get the neuron and GEMM operator number in CPU. \emph{Naghibijouybari et al.} show that side-channel effect in GPU platform can reveal the neuron numbers~\cite{CCS:2018:RenderGPU}. However, no direct evidence shows that how these statistics are useful to the attacking effectiveness. 
Targeting at the security in the edge(e.g.automotive), this work is the \emph{FIRST} to propose the NN model extraction methodology and experimentally conduct an \textbf{end-to-end attack} on an off-the-shelf \textbf{GPU platform} immune to \textbf{full system stack}(e.g.pytorch+cuDNN).


\section{Conclusion}\label{sec:conclusion}
The widespread use of neural network-based AI applications means that there is more incentive than ever before for attackers extract an accurate picture of inner functioning of their design. 
Through the acquisition of memory access events from bus snooping, layer sequence identification by the LSTM-CTC model, layer topology connection according to the memory access pattern, and layer dimension estimation under data volume constraints, we demonstrate one can accurately recover the a similar network architecture as the attack starting point.  
These reconstructed neural network  architectures present significant increase in attack success rate. 


\linespread{1}
\bibliographystyle{IEEEtran}
\bibliography{./main}

\begin{thebibliography}{10}
\providecommand{\url}[1]{#1}
\csname url@samestyle\endcsname
\providecommand{\newblock}{\relax}
\providecommand{\bibinfo}[2]{#2}
\providecommand{\BIBentrySTDinterwordspacing}{\spaceskip=0pt\relax}
\providecommand{\BIBentryALTinterwordstretchfactor}{4}
\providecommand{\BIBentryALTinterwordspacing}{\spaceskip=\fontdimen2\font plus
\BIBentryALTinterwordstretchfactor\fontdimen3\font minus
  \fontdimen4\font\relax}
\providecommand{\BIBforeignlanguage}[2]{{%
\expandafter\ifx\csname l@#1\endcsname\relax
\typeout{** WARNING: IEEEtran.bst: No hyphenation pattern has been}%
\typeout{** loaded for the language `#1'. Using the pattern for}%
\typeout{** the default language instead.}%
\else
\language=\csname l@#1\endcsname
\fi
#2}}
\providecommand{\BIBdecl}{\relax}
\BIBdecl

\bibitem{NIPS2012:Hinton}
\BIBentryALTinterwordspacing
A.~Krizhevsky, I.~Sutskever, and G.~E. Hinton, ``Imagenet classification with
  deep convolutional neural networks,'' in \emph{Proceedings of the 25th
  International Conference on Neural Information Processing Systems - Volume
  1}, ser. NIPS'12.\hskip 1em plus 0.5em minus 0.4em\relax USA: Curran
  Associates Inc., 2012, pp. 1097--1105. [Online]. Available:
  \url{http://dl.acm.org/citation.cfm?id=2999134.2999257}
\BIBentrySTDinterwordspacing

\bibitem{VGG2014}
\BIBentryALTinterwordspacing
K.~Simonyan and A.~Zisserman, ``Very deep convolutional networks for
  large-scale image recognition,'' \emph{CoRR}, vol. abs/1409.1556, 2014.
  [Online]. Available: \url{http://arxiv.org/abs/1409.1556}
\BIBentrySTDinterwordspacing

\bibitem{he2016deep}
K.~He, X.~Zhang, S.~Ren, and J.~Sun, ``Deep residual learning for image
  recognition,'' in \emph{Proceedings of the IEEE conference on computer vision
  and pattern recognition}, 2016, pp. 770--778.

\bibitem{xiong2017microsoft}
W.~Xiong, J.~Droppo, X.~Huang, F.~Seide, M.~Seltzer, A.~Stolcke, D.~Yu, and
  G.~Zweig, ``The microsoft 2016 conversational speech recognition system,'' in
  \emph{Acoustics, Speech and Signal Processing (ICASSP), 2017 IEEE
  International Conference on}.\hskip 1em plus 0.5em minus 0.4em\relax IEEE,
  2017, pp. 5255--5259.

\bibitem{NIPS2014:RNN}
\BIBentryALTinterwordspacing
I.~Sutskever, O.~Vinyals, and Q.~V. Le, ``Sequence to sequence learning with
  neural networks,'' in \emph{Advances in Neural Information Processing Systems
  27}, Z.~Ghahramani, M.~Welling, C.~Cortes, N.~D. Lawrence, and K.~Q.
  Weinberger, Eds.\hskip 1em plus 0.5em minus 0.4em\relax Curran Associates,
  Inc., 2014, pp. 3104--3112. [Online]. Available:
  \url{http://papers.nips.cc/paper/5346-sequence-to-sequence-learning-with-neural-networks.pdf}
\BIBentrySTDinterwordspacing

\bibitem{ICML2008:NLPNN}
\BIBentryALTinterwordspacing
R.~Collobert and J.~Weston, ``A unified architecture for natural language
  processing: Deep neural networks with multitask learning,'' in
  \emph{Proceedings of the 25th International Conference on Machine Learning},
  ser. ICML '08.\hskip 1em plus 0.5em minus 0.4em\relax New York, NY, USA: ACM,
  2008, pp. 160--167. [Online]. Available:
  \url{http://doi.acm.org/10.1145/1390156.1390177}
\BIBentrySTDinterwordspacing

\bibitem{vaswani2017attention}
A.~Vaswani, N.~Shazeer, N.~Parmar, J.~Uszkoreit, L.~Jones, A.~N. Gomez,
  {\L}.~Kaiser, and I.~Polosukhin, ``Attention is all you need,'' in
  \emph{Advances in Neural Information Processing Systems}, 2017, pp.
  6000--6010.

\bibitem{tractica_report}
T.~Report, ``Artificial intelligience market forecasts,'' 2016.

\bibitem{tesla_news}
Electrek, ``Elon musk clarifies tesla's plan for level 5 fully autonomous
  driving: 2 years away from sleeping in the car,'' 2017.

\bibitem{tesla_news2}
\BIBentryALTinterwordspacing
T.~Inc., ``Tesla autopilot: Full self-driving hardware on all cars,'' 2017.
  [Online]. Available: \url{https://www.tesla.com/autopilot}
\BIBentrySTDinterwordspacing

\bibitem{audi_nvidia}
TechCrunch, ``Nvidia is powering the world's first level 3 self-driving
  production car.'' 2017.

\bibitem{waymo1}
\BIBentryALTinterwordspacing
Waymo, ``Introducing waymo's suite of custom-build, self-driving hardware,''
  2017. [Online]. Available:
  \url{https://medium.com/waymo/introducing-waymos-suite-of-custom-built-self-driving-hardware-c47d1714563/}
\BIBentrySTDinterwordspacing

\bibitem{ucb}
D.~Amodei, C.~Olah, J.~Steinhardt, P.~Christiano, J.~Schulman, and D.~Man{\'e},
  ``Concrete problems in ai safety,'' \emph{arXiv preprint arXiv:1606.06565},
  2016.

\bibitem{atm_security}
\BIBentryALTinterwordspacing
C.~Middlehurst, ``{China unveils world's first facial recognition ATM},'' 2015.
  [Online]. Available:
  \url{https://www.telegraph.co.uk/news/worldnews/asia/china/11643314/China-unveils-worlds-first-facial-recognition-ATM.html}
\BIBentrySTDinterwordspacing

\bibitem{Ahmed_2015_CVPR}
E.~Ahmed, M.~Jones, and T.~K. Marks, ``An improved deep learning architecture
  for person re-identification,'' in \emph{The IEEE Conference on Computer
  Vision and Pattern Recognition (CVPR)}, June 2015.

\bibitem{NNsafetysummray:papernot2016}
N.~Papernot, P.~McDaniel, A.~Sinha, and M.~Wellman, ``Towards the science of
  security and privacy in machine learning,'' \emph{arXiv preprint
  arXiv:1611.03814}, 2016.

\bibitem{adversary:survey}
\BIBentryALTinterwordspacing
N.~Akhtar and A.~Mian, ``Threat of adversarial attacks on deep learning in
  computer vision: {A} survey,'' \emph{CoRR}, vol. abs/1801.00553, 2018.
  [Online]. Available: \url{http://arxiv.org/abs/1801.00553}
\BIBentrySTDinterwordspacing

\bibitem{ICLR17:Delving:Dawn}
\BIBentryALTinterwordspacing
Y.~Liu, X.~Chen, C.~Liu, and D.~Song, ``Delving into transferable adversarial
  examples and black-box attacks,'' \emph{ICLR}, vol. abs/1611.02770, 2017.
  [Online]. Available: \url{http://arxiv.org/abs/1611.02770}
\BIBentrySTDinterwordspacing

\bibitem{meta_learning}
\BIBentryALTinterwordspacing
B.~S. M.~F. Seong Joon~Oh, Max~Augustin, ``Towards reverse-engineering
  black-box neural networks,'' \emph{ICLR}, vol. abs/1605.07277, 2018.
  [Online]. Available: \url{https://arxiv.org/abs/1711.01768}
\BIBentrySTDinterwordspacing

\bibitem{fgsm}
I.~J. Goodfellow, J.~Shlens, and C.~Szegedy, ``Explaining and harnessing
  adversarial examples,'' \emph{Proceedings of the International Conference on
  Learning Representations}, 2015.

\bibitem{miFGSM}
\BIBentryALTinterwordspacing
C.~Szegedy, V.~Vanhoucke, S.~Ioffe, J.~Shlens, and Z.~Wojna, ``Rethinking the
  inception architecture for computer vision,'' \emph{CoRR}, vol.
  abs/1512.00567, 2015. [Online]. Available:
  \url{http://arxiv.org/abs/1512.00567}
\BIBentrySTDinterwordspacing

\bibitem{densenet}
G.~Huang, Z.~Liu, L.~Van Der~Maaten, and K.~Q. Weinberger, ``Densely connected
  convolutional networks.'' in \emph{CVPR 2017}.

\bibitem{resnet}
K.~He, X.~Zhang, S.~Ren, and J.~Sun, ``Deep residual learning for image
  recognition,'' in \emph{Proceedings of the IEEE conference on computer vision
  and pattern recognition}, 2016, pp. 770--778.

\bibitem{DAC18:ReverseNN}
\BIBentryALTinterwordspacing
W.~Hua, Z.~Zhang, and G.~E. Suh, ``Reverse engineering convolutional neural
  networks through side-channel information leaks,'' in \emph{Proceedings of
  the 55th Annual Design Automation Conference}, ser. DAC '18.\hskip 1em plus
  0.5em minus 0.4em\relax New York, NY, USA: ACM, 2018, pp. 4:1--4:6. [Online].
  Available: \url{http://doi.acm.org/10.1145/3195970.3196105}
\BIBentrySTDinterwordspacing

\bibitem{stealing_para}
\BIBentryALTinterwordspacing
F.~Tram\`{e}r, F.~Zhang, A.~Juels, M.~K. Reiter, and T.~Ristenpart, ``Stealing
  machine learning models via prediction apis,'' in \emph{Proceedings of the
  25th USENIX Conference on Security Symposium}, ser. SEC'16.\hskip 1em plus
  0.5em minus 0.4em\relax Berkeley, CA, USA: USENIX Association, 2016, pp.
  601--618. [Online]. Available:
  \url{http://dl.acm.org/citation.cfm?id=3241094.3241142}
\BIBentrySTDinterwordspacing

\bibitem{stealing_hyperpara}
\BIBentryALTinterwordspacing
B.~Wang and N.~Z. Gong, ``Stealing hyperparameters in machine learning,''
  \emph{CoRR}, vol. abs/1802.05351, 2018. [Online]. Available:
  \url{http://arxiv.org/abs/1802.05351}
\BIBentrySTDinterwordspacing

\bibitem{physical_attack:2002}
\BIBentryALTinterwordspacing
A.~Huang, ``Keeping secrets in hardware: The microsoft xbox\&\#153; case
  study,'' in \emph{Revised Papers from the 4th International Workshop on
  Cryptographic Hardware and Embedded Systems}, ser. CHES '02.\hskip 1em plus
  0.5em minus 0.4em\relax London, UK, UK: Springer-Verlag, 2003, pp. 213--227.
  [Online]. Available: \url{http://dl.acm.org/citation.cfm?id=648255.752707}
\BIBentrySTDinterwordspacing

\bibitem{DMA_Attack:12}
\BIBentryALTinterwordspacing
E.-O. Blass and W.~Robertson, ``Tresor-hunt: Attacking cpu-bound encryption,''
  in \emph{Proceedings of the 28th Annual Computer Security Applications
  Conference}, ser. ACSAC '12.\hskip 1em plus 0.5em minus 0.4em\relax New York,
  NY, USA: ACM, 2012, pp. 71--78. [Online]. Available:
  \url{http://doi.acm.org/10.1145/2420950.2420961}
\BIBentrySTDinterwordspacing

\bibitem{HMTT}
\BIBentryALTinterwordspacing
Y.~Huang, L.~Chen, Z.~Cui, Y.~Ruan, Y.~Bao, M.~Chen, and N.~Sun, ``Hmtt: A
  hybrid hardware/software tracing system for bridging the dram access trace's
  semantic gap,'' \emph{ACM Trans. Archit. Code Optim.}, vol.~11, no.~1, pp.
  7:1--7:25, Feb. 2014. [Online]. Available:
  \url{http://doi.acm.org/10.1145/2579668}
\BIBentrySTDinterwordspacing

\bibitem{CDMA}
M.~Rhu, M.~O'Connor, N.~Chatterjee, J.~Pool, Y.~Kwon, and S.~W. Keckler,
  ``Compressing dma engine: Leveraging activation sparsity for training deep
  neural networks,'' in \emph{2018 IEEE International Symposium on High
  Performance Computer Architecture (HPCA)}, Feb 2018, pp. 78--91.

\bibitem{GPUkernel:13}
\BIBentryALTinterwordspacing
D.~Lustig and M.~Martonosi, ``Reducing gpu offload latency via fine-grained
  cpu-gpu synchronization,'' in \emph{Proceedings of the 2013 IEEE 19th
  International Symposium on High Performance Computer Architecture (HPCA)},
  ser. HPCA '13.\hskip 1em plus 0.5em minus 0.4em\relax Washington, DC, USA:
  IEEE Computer Society, 2013, pp. 354--365. [Online]. Available:
  \url{http://dx.doi.org/10.1109/HPCA.2013.6522332}
\BIBentrySTDinterwordspacing

\bibitem{cuDNN}
\BIBentryALTinterwordspacing
Nvidia., ``Nvidia cudnn gpu accelerated deep learning,'' 2017. [Online].
  Available: \url{https://developer.nvidia.com/cudnn}
\BIBentrySTDinterwordspacing

\bibitem{ctc}
\BIBentryALTinterwordspacing
A.~Graves, S.~Fern\'{a}ndez, F.~Gomez, and J.~Schmidhuber, ``Connectionist
  temporal classification: Labelling unsegmented sequence data with recurrent
  neural networks,'' in \emph{Proceedings of the 23rd International Conference
  on Machine Learning}, ser. ICML '06.\hskip 1em plus 0.5em minus 0.4em\relax
  New York, NY, USA: ACM, 2006, pp. 369--376. [Online]. Available:
  \url{http://doi.acm.org/10.1145/1143844.1143891}
\BIBentrySTDinterwordspacing

\bibitem{GoogleNet}
C.~Szegedy, S.~Ioffe, V.~Vanhoucke, and A.~A. Alemi, ``Inception-v4,
  inception-resnet and the impact of residual connections on learning.'' in
  \emph{AAAI}, 2017, pp. 4278--4284.

\bibitem{polynet}
\BIBentryALTinterwordspacing
X.~Zhang, Z.~Li, C.~C. Loy, and D.~Lin, ``Polynet: {A} pursuit of structural
  diversity in very deep networks,'' \emph{CoRR}, vol. abs/1611.05725, 2016.
  [Online]. Available: \url{http://arxiv.org/abs/1611.05725}
\BIBentrySTDinterwordspacing

\bibitem{nasnet}
\BIBentryALTinterwordspacing
B.~Zoph, V.~Vasudevan, J.~Shlens, and Q.~V. Le, ``Learning transferable
  architectures for scalable image recognition,'' \emph{CoRR}, vol.
  abs/1707.07012, 2017. [Online]. Available:
  \url{http://arxiv.org/abs/1707.07012}
\BIBentrySTDinterwordspacing

\bibitem{BENCHIP}
J.-H. Tao, Z.-D. Du, Q.~Guo, H.-Y. Lan, L.~Zhang, S.-Y. Zhou, C.~Liu, H.-F.
  Liu, S.~Tang, and A.~Rush, ``Benchip: Benchmarking intelligence processors,''
  \emph{arXiv preprint arXiv:1710.08315}, 2017.

\bibitem{EndtoEnd_SR}
\BIBentryALTinterwordspacing
A.~Graves and N.~Jaitly, ``Towards end-to-end speech recognition with recurrent
  neural networks,'' in \emph{Proceedings of the 31st International Conference
  on International Conference on Machine Learning - Volume 32}, ser.
  ICML'14.\hskip 1em plus 0.5em minus 0.4em\relax JMLR.org, 2014, pp.
  II--1764--II--1772. [Online]. Available:
  \url{http://dl.acm.org/citation.cfm?id=3044805.3045089}
\BIBentrySTDinterwordspacing

\bibitem{intriguing:szegedy2013}
C.~Szegedy, W.~Zaremba, I.~Sutskever, J.~Bruna, D.~Erhan, I.~Goodfellow, and
  R.~Fergus, ``Intriguing properties of neural networks,'' \emph{arXiv preprint
  arXiv:1312.6199}, 2013.

\bibitem{targetRegressionAttack:AAAI}
\BIBentryALTinterwordspacing
S.~Alfeld, X.~Zhu, and P.~Barford, ``Data poisoning attacks against
  autoregressive models,'' in \emph{Proceedings of the Thirtieth AAAI
  Conference on Artificial Intelligence}, ser. AAAI'16.\hskip 1em plus 0.5em
  minus 0.4em\relax AAAI Press, 2016, pp. 1452--1458. [Online]. Available:
  \url{http://dl.acm.org/citation.cfm?id=3016100.3016102}
\BIBentrySTDinterwordspacing

\bibitem{physicalMLattack:2016}
A.~Kurakin, I.~Goodfellow, and S.~Bengio, ``Adversarial examples in the
  physical world,'' \emph{arXiv preprint arXiv:1607.02533}, 2016.

\bibitem{Papernot:2017:blackbox}
\BIBentryALTinterwordspacing
N.~Papernot, P.~McDaniel, I.~Goodfellow, S.~Jha, Z.~B. Celik, and A.~Swami,
  ``Practical black-box attacks against machine learning,'' in
  \emph{Proceedings of the 2017 ACM on Asia Conference on Computer and
  Communications Security}, ser. ASIA CCS '17.\hskip 1em plus 0.5em minus
  0.4em\relax New York, NY, USA: ACM, 2017, pp. 506--519. [Online]. Available:
  \url{http://doi.acm.org/10.1145/3052973.3053009}
\BIBentrySTDinterwordspacing

\bibitem{axiv:papernot:transferability}
\BIBentryALTinterwordspacing
N.~Papernot, P.~D. McDaniel, and I.~J. Goodfellow, ``Transferability in machine
  learning: from phenomena to black-box attacks using adversarial samples,''
  \emph{CoRR}, vol. abs/1605.07277, 2016. [Online]. Available:
  \url{http://arxiv.org/abs/1605.07277}
\BIBentrySTDinterwordspacing

\bibitem{NVIDIA_K40}
NVIDIA, ``Nvidia tesla k40 active gpu accelerator,''
  \url{http://www.pny.com/nvidia-tesla-k40-active-gpu-accelerator}, 2016.

\bibitem{pytorch}
\BIBentryALTinterwordspacing
PyTorch., ``Pytorch tutorials.'' [Online]. Available:
  \url{http://pytorch.org/tutorials/}
\BIBentrySTDinterwordspacing

\bibitem{cuda}
N.~Wilt, \emph{The cuda handbook: A comprehensive guide to gpu
  programming}.\hskip 1em plus 0.5em minus 0.4em\relax Pearson Education, 2013.

\bibitem{nvprof}
\BIBentryALTinterwordspacing
Nvidia., ``Cuda toolkit documentation.'' [Online]. Available:
  \url{http://docs.nvidia.com/cuda/profiler-users-guide/index.html}
\BIBentrySTDinterwordspacing

\bibitem{adam}
\BIBentryALTinterwordspacing
D.~P. Kingma and J.~Ba, ``Adam: A method for stochastic optimization.''
  \emph{CoRR}, vol. abs/1412.6980, 2014. [Online]. Available:
  \url{http://dblp.uni-trier.de/db/journals/corr/corr1412.html#KingmaB14}
\BIBentrySTDinterwordspacing

\bibitem{squeezenet}
F.~N. Iandola, S.~Han, M.~W. Moskewicz, K.~Ashraf, W.~J. Dally, and K.~Keutzer,
  ``Squeezenet: Alexnet-level accuracy with 50x fewer parameters and< 0.5 mb
  model size,'' \emph{arXiv preprint arXiv:1602.07360}, 2016.

\bibitem{deng2009imagenet}
J.~Deng, W.~Dong, R.~Socher, L.-J. Li, K.~Li, and L.~Fei-Fei, ``Imagenet: A
  large-scale hierarchical image database,'' in \emph{Computer Vision and
  Pattern Recognition, 2009. CVPR 2009. IEEE Conference on}.\hskip 1em plus
  0.5em minus 0.4em\relax IEEE, 2009, pp. 248--255.

\bibitem{path_oram:2013}
\BIBentryALTinterwordspacing
E.~Stefanov, M.~van Dijk, E.~Shi, C.~Fletcher, L.~Ren, X.~Yu, and S.~Devadas,
  ``Path oram: An extremely simple oblivious ram protocol,'' in
  \emph{Proceedings of the 2013 ACM SIGSAC Conference on Computer \&\#38;
  Communications Security}, ser. CCS '13.\hskip 1em plus 0.5em minus
  0.4em\relax New York, NY, USA: ACM, 2013, pp. 299--310. [Online]. Available:
  \url{http://doi.acm.org/10.1145/2508859.2516660}
\BIBentrySTDinterwordspacing

\bibitem{Liu:ORAM}
\BIBentryALTinterwordspacing
C.~Liu, A.~Harris, M.~Maas, M.~Hicks, M.~Tiwari, and E.~Shi, ``Ghostrider: A
  hardware-software system for memory trace oblivious computation,'' in
  \emph{Proceedings of the Twentieth International Conference on Architectural
  Support for Programming Languages and Operating Systems}, ser. ASPLOS
  '15.\hskip 1em plus 0.5em minus 0.4em\relax New York, NY, USA: ACM, 2015, pp.
  87--101. [Online]. Available:
  \url{http://doi.acm.org/10.1145/2694344.2694385}
\BIBentrySTDinterwordspacing

\bibitem{Liu:2013:ORAM}
\BIBentryALTinterwordspacing
C.~Liu, M.~Hicks, and E.~Shi, ``Memory trace oblivious program execution,'' in
  \emph{Proceedings of the 2013 IEEE 26th Computer Security Foundations
  Symposium}, ser. CSF '13.\hskip 1em plus 0.5em minus 0.4em\relax Washington,
  DC, USA: IEEE Computer Society, 2013, pp. 51--65. [Online]. Available:
  \url{https://doi.org/10.1109/CSF.2013.11}
\BIBentrySTDinterwordspacing

\bibitem{tvm}
\BIBentryALTinterwordspacing
T.~Chen, T.~Moreau, Z.~Jiang, H.~Shen, E.~Q. Yan, L.~Wang, Y.~Hu, L.~Ceze,
  C.~Guestrin, and A.~Krishnamurthy, ``{TVM:} end-to-end optimization stack for
  deep learning,'' \emph{CoRR}, vol. abs/1802.04799, 2018. [Online]. Available:
  \url{http://arxiv.org/abs/1802.04799}
\BIBentrySTDinterwordspacing

\bibitem{mitAttackSummary:18}
N.~Akhtar and A.~Mian, ``Threat of adversarial attacks on deep learning in
  computer vision: A survey,'' \emph{arXiv preprint arXiv:1801.00553}, 2018.

\bibitem{CoRR15:Papernot}
\BIBentryALTinterwordspacing
S.~J. M. F. Z. B. C. A.~S. Nicolas~Papernot, Patrick D.~McDaniel, ``The
  limitations of deep learning in adversarial settings,'' \emph{CoRR}, vol.
  abs/1511.07528, 2015. [Online]. Available:
  \url{http://arxiv.org/abs/1511.07528}
\BIBentrySTDinterwordspacing

\bibitem{carlini2017towards}
N.~Carlini and D.~Wagner, ``Towards evaluating the robustness of neural
  networks,'' in \emph{2017 IEEE Symposium on Security and Privacy (SP)}.\hskip
  1em plus 0.5em minus 0.4em\relax IEEE, 2017, pp. 39--57.

\bibitem{baluja2017adversarial}
S.~Baluja and I.~Fischer, ``Adversarial transformation networks: Learning to
  generate adversarial examples,'' \emph{arXiv preprint arXiv:1703.09387},
  2017.

\bibitem{das2011differential}
S.~Das and P.~N. Suganthan, ``Differential evolution: a survey of the
  state-of-the-art,'' \emph{IEEE transactions on evolutionary computation},
  vol.~15, no.~1, pp. 4--31, 2011.

\bibitem{cisse2017houdini}
M.~Cisse, Y.~Adi, N.~Neverova, and J.~Keshet, ``Houdini: Fooling deep
  structured prediction models,'' \emph{arXiv preprint arXiv:1707.05373}, 2017.

\bibitem{sarkar2017upset}
S.~Sarkar, A.~Bansal, U.~Mahbub, and R.~Chellappa, ``Upset and angri: Breaking
  high performance image classifiers,'' \emph{arXiv preprint arXiv:1707.01159},
  2017.

\bibitem{deepfool:cpvr2016}
S.~M. Moosavi~Dezfooli, A.~Fawzi, and P.~Frossard, ``Deepfool: a simple and
  accurate method to fool deep neural networks,'' in \emph{Proceedings of 2016
  IEEE Conference on Computer Vision and Pattern Recognition (CVPR)}, no.
  EPFL-CONF-218057, 2016.

\bibitem{moosavi2017universal}
S.-M. Moosavi-Dezfooli, A.~Fawzi, O.~Fawzi, and P.~Frossard, ``Universal
  adversarial perturbations,'' \emph{arXiv preprint}, 2017.

\bibitem{su2017one}
J.~Su, D.~V. Vargas, and S.~Kouichi, ``One pixel attack for fooling deep neural
  networks,'' \emph{arXiv preprint arXiv:1710.08864}, 2017.

\bibitem{mopuri2017fast}
K.~R. Mopuri, U.~Garg, and R.~V. Babu, ``Fast feature fool: A data independent
  approach to universal adversarial perturbations,'' \emph{arXiv preprint
  arXiv:1707.05572}, 2017.

\bibitem{KDD2013:evasion}
\BIBentryALTinterwordspacing
B.~Biggio, I.~Corona, D.~Maiorca, B.~Nelson, N.~\v{S}rndi\'{c}, P.~Laskov,
  G.~Giacinto, and F.~Roli, ``Evasion attacks against machine learning at test
  time,'' in \emph{Proceedings of the 2013th European Conference on Machine
  Learning and Knowledge Discovery in Databases - Volume Part III}, ser.
  ECMLPKDD'13.\hskip 1em plus 0.5em minus 0.4em\relax Berlin, Heidelberg:
  Springer-Verlag, 2013, pp. 387--402. [Online]. Available:
  \url{https://doi.org/10.1007/978-3-642-40994-3_25}
\BIBentrySTDinterwordspacing

\bibitem{realFaceDetactionAttack:2016}
M.~Sharif, S.~Bhagavatula, L.~Bauer, and M.~K. Reiter, ``Accessorize to a
  crime: Real and stealthy attacks on state-of-the-art face recognition,'' in
  \emph{Proceedings of the 2016 ACM SIGSAC Conference on Computer and
  Communications Security}.\hskip 1em plus 0.5em minus 0.4em\relax ACM, 2016,
  pp. 1528--1540.

\bibitem{malwareClassification:2016}
M.~Backes, P.~Manoharan, K.~Grosse, and N.~Papernot, ``Adversarial
  perturbations against deep neural networks for malware classification,''
  \emph{The Computing Research Repository (CoRR)}, 2016.

\bibitem{ateniese2013hacking}
G.~Ateniese, G.~Felici, L.~V. Mancini, A.~Spognardi, A.~Villani, and D.~Vitali,
  ``Hacking smart machines with smarter ones: How to extract meaningful data
  from machine learning classifiers,'' \emph{arXiv preprint arXiv:1306.4447},
  2013.

\bibitem{sabour2015adversarial}
S.~Sabour, Y.~Cao, F.~Faghri, and D.~J. Fleet, ``Adversarial manipulation of
  deep representations,'' \emph{arXiv preprint arXiv:1511.05122}, 2015.

\bibitem{rozsa2016adversarial}
A.~Rozsa, E.~M. Rudd, and T.~E. Boult, ``Adversarial diversity and hard
  positive generation,'' in \emph{Proceedings of the IEEE Conference on
  Computer Vision and Pattern Recognition Workshops}, 2016, pp. 25--32.

\bibitem{PracticalEvasion}
N.~rndic and P.~Laskov, ``Practical evasion of a learning-based classifier: A
  case study,'' in \emph{2014 IEEE Symposium on Security and Privacy}, May
  2014, pp. 197--211.

\bibitem{shokri2017membership}
R.~Shokri, M.~Stronati, C.~Song, and V.~Shmatikov, ``Membership inference
  attacks against machine learning models,'' in \emph{Security and Privacy
  (SP), 2017 IEEE Symposium on}.\hskip 1em plus 0.5em minus 0.4em\relax IEEE,
  2017, pp. 3--18.

\bibitem{xu2016automatically}
W.~Xu, Y.~Qi, and D.~Evans, ``Automatically evading classifiers,'' in
  \emph{Proceedings of the 2016 Network and Distributed Systems Symposium},
  2016.

\bibitem{fredrikson2014privacy}
M.~Fredrikson, E.~Lantz, S.~Jha, S.~Lin, D.~Page, and T.~Ristenpart, ``Privacy
  in pharmacogenetics: An end-to-end case study of personalized warfarin
  dosing.'' in \emph{USENIX Security Symposium}, 2014, pp. 17--32.

\bibitem{narodytska2016simple}
N.~Narodytska and S.~P. Kasiviswanathan, ``Simple black-box adversarial
  perturbations for deep networks,'' \emph{arXiv preprint arXiv:1612.06299},
  2016.

\bibitem{joon2017adversarial}
S.~Joon~Oh, M.~Fritz, and B.~Schiele, ``Adversarial image perturbation for
  privacy protection--a game theory perspective,'' in \emph{Proceedings of the
  IEEE International Conference on Computer Vision}, 2017, pp. 1482--1491.

\bibitem{dong2017boosting}
Y.~Dong, F.~Liao, T.~Pang, H.~Su, X.~Hu, J.~Li, and J.~Zhu, ``Boosting
  adversarial attacks with momentum,'' \emph{arXiv preprint arXiv:1710.06081},
  2017.

\bibitem{carlini2017ground}
N.~Carlini, G.~Katz, C.~Barrett, and D.~L. Dill, ``Ground-truth adversarial
  examples,'' \emph{arXiv preprint arXiv:1709.10207}, 2017.

\bibitem{ICCAD:Qiang}
\BIBentryALTinterwordspacing
L.~Wei, Y.~Liu, B.~Luo, Y.~Li, and Q.~Xu, ``I know what you see: Power
  side-channel attack on convolutional neural network accelerators,''
  \emph{CoRR}, vol. abs/1803.05847, 2018. [Online]. Available:
  \url{http://arxiv.org/abs/1803.05847}
\BIBentrySTDinterwordspacing

\bibitem{arXiv18:cacheTelpathy}
\BIBentryALTinterwordspacing
M.~Yan, C.~W. Fletcher, and J.~Torrellas, ``Cache telepathy: Leveraging shared
  resource attacks to learn {DNN} architectures,'' \emph{CoRR}, vol.
  abs/1808.04761, 2018. [Online]. Available:
  \url{http://arxiv.org/abs/1808.04761}
\BIBentrySTDinterwordspacing

\bibitem{CCS:2018:RenderGPU}
\BIBentryALTinterwordspacing
H.~Naghibijouybari, A.~Neupane, Z.~Qian, and N.~Abu-Ghazaleh, ``Rendered
  insecure: Gpu side channel attacks are practical,'' in \emph{Proceedings of
  the 2018 ACM SIGSAC Conference on Computer and Communications Security}, ser.
  CCS '18.\hskip 1em plus 0.5em minus 0.4em\relax New York, NY, USA: ACM, 2018,
  pp. 2139--2153. [Online]. Available:
  \url{http://doi.acm.org/10.1145/3243734.3243831}
\BIBentrySTDinterwordspacing

\end{thebibliography}

\end{document}